\documentclass[letterpaper,11pt]{article}
\pdfoutput=1 

\usepackage{jcappub} 

\usepackage[T1]{fontenc} 

\usepackage{amsmath}
\def\beq{\begin{equation}}
\def\eeq{\end{equation}}

\DeclareMathOperator{\csch}{csch}

\title{Reheating Dynamics in Inflationary Cosmology: Insights from $\alpha$-Attractor and $\alpha$-Starobinsky Models}

\author[]{Gabriel Germ\'an}
\affiliation[]{Instituto de Ciencias F\'{\i}sicas, Universidad Nacional
Aut\'onoma de M\'exico,\\ Av. Universidad s/n, Cuernavaca, Morelos, 62210, Mexico}

\emailAdd{gabriel@icf.unam.mx}

\abstract{
Reheating in inflationary cosmology is essential for understanding the early universe, influencing particle production, thermalization, and the primordial power spectrum. Crucial quantities defined during the reheating epoc, such as the equation of state parameter $\omega_{re}$, reheating temperature $T_{re}$, and the number of $e$-folds $N_{re}$, affect inflationary observables like the scalar spectral index $n_s$ and the tensor-to-scalar ratio $r$.
We analyze two classes of inflationary models: generalized $\alpha$-attractor models and the $\alpha$-Starobinsky generalization. These models, motivated by supergravity and string theory, exhibit attractor behavior, ensuring strong predictions and have been studied extensively before. An important novelty of this study, compared to previous works, is the inclusion of an analytical expression for the reheating temperature, $T_{\text{re}}$, which makes it a dynamical quantity. This is crucial for determining all the cosmological quantities analyzed in this work. Our results show a universal scaling behavior for a tightly bounded $T_{re}$ in both models. We believe this is the first time that $T_{re}$ is so closely determined. This work complements previous Bayesian and numerical studies by providing detailed numerical and analytical insights into the evolution of cosmological observables and reheating parameters, offering also constraints on inflationary models based on observational data.
}

\begin{document}
\maketitle
\flushbottom
\section{Introduction}\label{Intro}

The study of reheating in inflationary cosmology is critical for understanding the early universe and provides key insights into the dynamics and parameters of inflation (for reviews on inflation see e.g., \cite{Linde:1984ir}-\cite{Martin:2013tda}). Reheating  follows the end of inflation and is characterized by basic quantities such as the equation of state (EoS) parameter, $\omega_{re}$, the reheating temperature, $T_{re}$, and the duration of the reheating phase, typically expressed in terms of the number of $e$-folds, $N_{re}$ (for reviews on reheating see e.g., \cite{Bassett:2005xm}-\cite{Amin:2014eta}). These quantities have important implications for particle production, thermalization, and the primordial power spectrum. Specifically, the duration of reheating influences the abundance of particles, including dark matter, and affects the observable anisotropies in the cosmic microwave background (CMB).

By combining cosmological observations, such as those from the CMB, with theoretical models, constraints on the EoS parameter $\omega_{re}$ can be derived. These constraints, in turn, limit key inflationary observables, including the scalar spectral index $n_s$ and the tensor-to-scalar ratio $r$. Consistency relations between these observables provide a framework for testing the internal coherence of inflationary models and for guiding the selection of priors in Bayesian analyses.

In this paper, we focus on two well known classes of inflationary models: the generalized $\alpha$-attractor models 
\begin{equation}
    V(\phi) = V_0\,  \text{tanh}^2\left(\frac{\phi}{\sqrt{6\alpha}M_{Pl}}\right),
    \label{A_pot_0}
\end{equation}
and the $\alpha$-Starobinsky generalization
\begin{equation}
    \label{VS_0}
    V(\phi) = V_0 \left(1 - e^{-\sqrt{\frac{2}{3\alpha}}\frac{\phi}{M_{Pl}}}\right)^2,
\end{equation}
where $\alpha$ is a parameter related to the inflationary potential. (not to be confused with the running of the scalar spectral index, here denoted by  $n_{sk}$),  and $M_{Pl} \approx 2.44 \times 10^{18} \, \mathrm{GeV}$ is the reduced Planck mass. Generalized $\alpha$-attractor models are motivated by supergravity and string theory and exhibit connections to hyperbolic and logarithmic geometries in field space \cite{Ferrara:2013rsa}-\cite{Kallosh:2013tua}. These models are characterized by attractor behavior, ensuring robustness against variations in initial conditions and consistently reproducing observed values of $n_s$, $r$, and other inflationary observables.

The Starobinsky model \cite{Starobinsky:1980te}, a geometric extension of general relativity, also remains in excellent agreement with current observational data, including constraints on $n_s$ and $r$ from the Planck and BICEP/Keck collaborations~\cite{Akrami:2018odb}-\cite{Tristram:2021tvh}. Originally formulated in the Jordan frame, the Starobinsky model can be transformed into the Einstein frame via a conformal transformation, where it becomes equivalent to a single-field inflationary model with an asymptotically flat potential. This transformation naturally introduces couplings between the inflaton and Standard Model fields, providing a mechanism for reheating. Generalizations of the Starobinsky model, such as the $\alpha$-Starobinsky models, incorporate supergravity-based deformations of the inflaton potential~ \cite{Kallosh:2013yoa}, \cite{Ellis:2013nxa}.

The aim of this article is to complement previous Bayesian and numerical studies of inflationary models by providing a detailed analysis of the evolution of cosmological observables, the number of $e$-folds, and the reheating temperature as functions of $\omega_{re}$, and the scalar spectral index, $n_s$. 
To achieve this, we numerically solve an equation that links the inflationary phase to the reheating phase over a range of values for the model parameter $\alpha$ and  $\omega_{re}$. This allows us to generate a set of plots that illustrate the behavior of the cosmological observables and quantities mentioned above.
The novel aspect of this study, compared to previous works, is the derivation, from inflaton decay into light scalar fields, of an explicit analytical expression for a dynamical reheating temperature, $T_{re}$, which plays a key role in determining all the cosmological quantities discussed in this analysis. This enables us to study $T_{re}$ as a function of the number of $e$-folds during inflation, the spectral index, and other related parameters. This expression for $T_{re}$ enters the main equation connecting inflation with reheating, and consequently, all the plots presented here reflect the evolution of $T_{re}$. We also show that a large-$N_k$ expansion of $T_{re}$ for the $\alpha$-attractor and the $\alpha$-Starobinsky models yields the same leading term, scaling as a power of $1/N_k^{3/2}$, in agreement with similar universal results for these classes of models. Our results show  a tightly bounded reheating temperature in both models.

The article is organized as follows: In Section 2, we review the theoretical background connecting inflationary models and reheating, including a discussion of inflaton decay and its role in determining $T_{re}$ as a function of the inflaton field. In Section 3, we provide detailed plots of observables, the number of $e$-folds, and the reheating temperature as functions of the EoS parameter $\omega_{re}$, and the scalar spectral index $n_s$ for the $\alpha$-attractor model. In Section 4, we present similar results for the $\alpha$-Starobinsky model. Finally, Section 5 contains our conclusions and prospects for future work.
\section {The general approach}\label{general}In the following, we review general aspects of reheating, which will be applied in the subsequent analysis of two specific models. In subsection \ref{reheating}, we present the general formalism assuming a constant, effective equation of state parameter, $\omega_{re}$, during reheating. This culminates in equations (\ref{omega1}) and (\ref{Nk2}), which are equivalent and connect the inflationary phase to the reheating phase. In subsection \ref{temperature}, we derive an original analytic expression for the reheating temperature, eq.~(\ref{TreGeneral}), at the end of reheating. This formula allows us to compute the range of the reheating temperature, particularly for the case where the inflaton decays into light scalar fields, given a complete model of inflation.
Interestingly, the highest value of $T_{re}$ is very close to the upper bound necessary to avoid the overproduction of gravitinos during reheating, within the framework of supergravity theories. We also show that, for large $N_k$ (the number of $e$-folds during inflation), $T_{re}$ exhibits the same leading term, $M_{Pl}/N_k^{3/2}$, independent of $\alpha$, for both the $\alpha$-attractor and $\alpha$-Starobinsky models. This is consistent with the universal behavior observed in this class of models.
\subsection {Introducing reheating}\label{reheating}
Most of the equations presented below are well-known  \cite{Liddle:1994dx}-\cite{German:2023yer}, but are included here for ease of reference and clarity of exposition.
The number of $e$-folds during reheating, denoted by $N_{re}$, is defined as
\begin{equation}
\label{Nre1}
N_{re} \equiv \ln\left(\frac{a_{re}}{a_{end}}\right),
\end{equation}
where $a_{re}$ and $a_{end}$ are the scale factors at the end of reheating and inflation, respectively. This can alternatively be expressed as
\begin{equation}
\label{Nre2}
N_{re} = \frac{1}{3\left(1 + \omega_{re}\right)} \ln\left(\frac{\rho_{end}}{\rho_{re}}\right),
\end{equation}
where $\omega_{re}$ is the effective equation of state during reheating (assumed constant), and $\rho_{end}$ and $\rho_{re}$ are the energy densities at the end of inflation and at the end of reheating, respectively. The energy density at the end of inflation is given by
\begin{equation}
\label{roend}
\rho_{end} = \frac{3}{2} V_{end} = \frac{9}{2} \frac{V_{end}}{V_k} H_k^2 M_{Pl}^2,
\end{equation}
where $V_{end}$ is the potential at the end of inflation, $V_k$ is the potential at horizon crossing, $H_k$ is the Hubble parameter at horizon crossing. From the amplitude of scalar perturbations, we have
\begin{equation}
\label{Hk}
H_k = \pi \sqrt{\frac{A_s r}{2}} M_{Pl},
\end{equation}
where $A_s$ is the amplitude of the scalar power spectrum, and $r$ is the tensor-to-scalar ratio. The energy density at the end of reheating is given by
\begin{equation}
\label{rore1}
\rho_{re} = \frac{\pi^2 g_{re}}{30} T_{re}^4,
\end{equation}
where $T_{re}$ is the reheating temperature, and $g_{re}$ is the effective number of relativistic degrees of freedom at that time.

The reheating temperature $T_{re}$ can be obtained from the conservation of entropy after reheating. The number of $e$-folds during radiation domination, $N_{rd}$, is 
\begin{equation}
\label{IONrd}
N_{rd} \equiv \ln\left(\frac{a_{eq}}{a_r}\right) = \ln\left(\frac{a_{eq} T_{re}}{\left(\frac{43}{11 g_{s,re}}\right)^{1/3} a_0 T_0}\right),
\end{equation}
where $a_{eq}$ is the scale factor at radiation-matter equality, $g_{s,re}$ is the number of entropy degrees of freedom at reheating, and $a_0$ and $T_0$ are the current scale factor and temperature, respectively.

Next, we compute the total expansion from when a mode of comoving wavelength $k$ exited the horizon during inflation to the time of radiation-matter equality, denoted by $N_{keq}$
\begin{equation}
\label{Nkeq1}
N_{keq} \equiv \ln\left(\frac{a_{eq}}{a_k}\right) = \ln\left(\frac{a_{end}}{a_k}\right) + \ln\left(\frac{a_{re}}{a_{end}}\right) + \ln\left(\frac{a_{eq}}{a_{re}}\right) = N_{k} + N_{re} + N_{rd},
\end{equation}
where $N_k$ is the number of $e$-folds from horizon crossing to the end of inflation. Another expression for $N_{keq}$ is \cite{German:2020iwg}
\begin{equation}
\label{Nkeq2}
N_{keq} \equiv \ln\left(\frac{a_{eq}}{a_k}\right) = \ln\left(\frac{a_{eq} H_k}{k_p}\right) = \ln\left(\frac{a_{eq} \pi \sqrt{A_s r}}{\sqrt{2} k_p}\right),
\end{equation}
where $k_p$ is the pivot scale. Using eq.~(\ref{Nkeq1}) to express $N_{keq}$, we can eliminate $N_{re}$ from eq.~(\ref{Nre2})
\begin{equation}
\label{Nk1}
N_k = N_{keq} - N_{rd} - \frac{1}{3\left(1 + \omega_{re}\right)} \ln\left(\frac{\rho_{end}}{\rho_{re}}\right),
\end{equation}
so that $\omega_{re}$ appears explicitly only in the shown term. From this expression, we can isolate $\omega_{re}$ as \cite{Garcia:2023tkk}
\begin{equation}
\label{omega1}
\omega_{re}= -1-\frac{\ln\left(\rho_{end}/\rho_{re}\right)}{3\left(N_k - N_{keq} + N_{rd}\right)}.
\end{equation}
This formula will be useful for constraining the spectral index $n_s$ and the tensor-to-scalar ratio $r$, given a specific range of $\omega_{re}$ values and viceversa, as we will show in the following sections.

An alternative way of presenting eq.~(\ref{omega1}) can be derived as follows \cite{German:2023yer}. In general, the potential can be written as $V(\phi) = V_0 f(\phi)$, where $V_0$ is an overall energy scale factor, and $f(\phi)$ contains the dependence on the inflaton field $\phi$. Thus, eq.~(\ref{Nk1}) can be explicitly written as
\begin{equation}
\label{Nk2}
\begin{split}
N_{k} = & \ln\left(\frac{2}{k_p} \left(\frac{43}{11 g_{s,re}}\right)^{1/3} \pi \sqrt{A_s} a_0 T_0 \right) 
+ \frac{1}{3(1 + \omega_{re})} \ln\left(\frac{g_{re}}{540 A_s}\right) + \frac{1 - 3 \omega_{re}}{3(1 + \omega_{re})} \ln \left(\frac{T_{re}}{M_{Pl}}\right)\\
&  + \frac{1}{3(1 + \omega_{re})} \ln\left(\frac{\left(M_{Pl} f^{\prime}(\phi_k)\right)^{1 + 3 \omega_{re}}}{f(\phi_k)^{3 \omega_{re}} f(\phi_{end})}\right),
\end{split}
\end{equation}
where $f'(\phi_k)$ is the derivative of $f(\phi)$ with respect to $\phi$, evaluated at $\phi = \phi_k$ (the value of the inflaton at horizon crossing). Eq.~(\ref{Nk2}) can be solved for $N_k$ once $\phi_k$ is expressed in terms of $N_k$, or $\phi_k$ can be determined directly once the model is specified. This will be done in the next sections thus, it is convenient to have $T_{re}$ expressed in terms of $\phi_k$ as done in subsection \ref{temperature} below.

The right-hand side of eq.~(\ref{Nk2}) consists of four distinct terms: a constant term depending on cosmological parameters such as $A_s$, $g_{s,re}$, $T_0$, and the pivot scale $k_p$; terms involving the equation of state parameter $\omega_{re}$ and the reheating temperature $T_{re}$; and a final term that captures the model dependence via the normalized inflationary potential $f(\phi)$ and its derivative. The left-hand side ($N_k$) is expressed in terms of the value of $\phi$ at the end of inflation, $\phi_{end}$, as well as the model parameters and $\phi_k$, the value of the inflaton at horizon crossing. We will see that $T_{re}$ can be expressed in terms of the parameter of the model $\alpha$, and $\phi_k$. Thus, in what follows, we treat the parameters $\alpha$, and $\omega_{re}$ as free variables. For a given inflationary model and a pair ($\alpha$, $\omega_{re}$), we compute $\phi_k$ from eq.~(\ref{Nk2}), and from this, all observables and cosmological quantities of interest can be derived straightforwardly. 
\subsection {The thermalization temperature}\label{temperature}

We now turn our attention to the reheating temperature at the end of the reheating phase, or thermalization temperature. If the inflaton couples more strongly to scalar fields than to other particles, decays into scalars will dominate. For simplicity, assuming this to be the case, the decay rate into light scalar fields in perturbation theory, when the coupling is primarily due to gravitational effects, can be approximated as
\begin{equation}
\label{Decay}
\Gamma_{\phi}\approx \frac{m_{\phi}^3}{192\pi M_{Pl}^2},
\end{equation}
where $m_{\phi}$ is the inflaton mass, given by
\begin{equation}
\label{mass}
m_{\phi}^2=V^{\prime \prime}(\phi_0)\approx \beta V_0,
\end{equation}
with $\beta$ being a dimensionful model-dependent parameter to be determined for each specific case, and $\phi_0$ the value of the inflaton at the minimum of the potential. Using the amplitude of scalar perturbations at horizon crossing
\begin{equation}
\label{A1}
A_s(k) =\frac{1}{24\pi^2} \frac{V(\phi_k)}{\epsilon_k\, M_{Pl}^{4}},
\end{equation}
we can express the potential as
\begin{equation}
\label{V01}
V(\phi_k)\approx V_0\approx  \frac{3}{2}\pi^2 A_s r M_{Pl}^4.
\end{equation}
At the end of reheating, the energy density of the inflaton can be approximated as
\begin{equation}
\label{rore2}
\rho_{re}\approx 3 H_{re}^2M_{Pl}^2 \approx 3 \Gamma_{\phi}^2M_{Pl}^2.
\end{equation}
Reheating concludes when the energy densities of radiation and inflaton decay become comparable. At this stage, the reheating temperature can be determined by
\begin{equation}
\label{rorad}
\rho_{re}=\frac{\pi^2 g_{re}}{30}T_{re}^4,
\end{equation}
where $g_{re}$ is the effective number of relativistic degrees of freedom. Combining these expressions gives
\begin{equation}
\label{Tre}
T_{re}\approx \frac{1}{\sqrt{8}}\left(\frac{135 A_s^3 \pi^2 M_{Pl}^{10} r^3 \beta^3}{256 g_{re}}\right)^{1/4}.
\end{equation}

As particular examples, let us now consider two models, which will be discussed in detail in the following sections. For the $\alpha$-attractor model of Section \ref{Attractor}, we have $\beta_A = \frac{1}{3\alpha M_{Pl}^2}$, and the universal asymptotic behavior of the tensor-to-scalar ratio in this class of models is $r \approx \frac{12\alpha}{N_k^2}$. Thus,
\begin{equation}
\label{TreA}
T_{re}(A) \approx \left(\frac{135 A_s^3 \pi^2}{256 g_{re}}\right)^{1/4} \frac{M_{Pl}}{N_k^{3/2}}.
\end{equation}
For the $\alpha$-Starobinsky model of Section \ref{Staro}, $\beta_S = \frac{4}{3\alpha M_{Pl}^2}$, and
\begin{equation}
\label{TreS}
T_{re}(S) \approx \sqrt{8} \left(\frac{135 A_s^3 \pi^2}{256 g_{re}}\right)^{1/4} \frac{M_{Pl}}{N_k^{3/2}}.
\end{equation}
Thus, $T_{re}(S) \approx 2.83 T_{re}(A)$, note that both results are $\alpha$-independent.

\begin{figure*}[h!]
\begin{center}
\includegraphics[width=4.5 in]{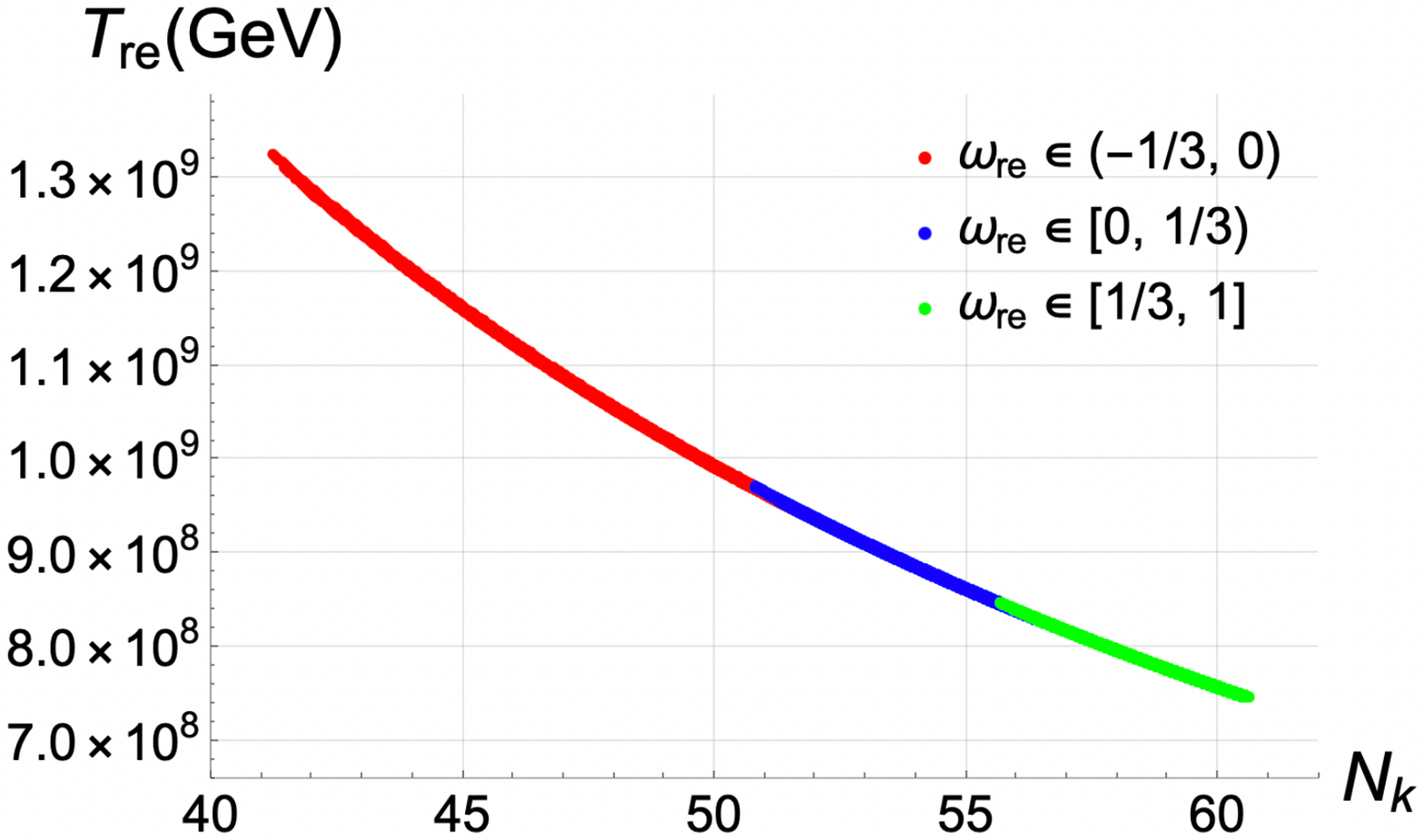}
\caption{The figure shows the number of $e$-folds during inflation, $N_k$, from eq.~(\ref{NkA1}), plotted against the reheating temperature, $T_{re}$, from eq.~(\ref{TreA1}), for the $\alpha$-attractor model defined by eq.~(\ref{A_pot}). The plot is generated by first solving the full eq.~(\ref{Nk2}) for $\phi_k$, with the plotted quantities derived from their respective defining equations. The EoS interval $-1/3 < \omega_{re} < 1$ is divided into three regions, as specified. The parameter $\alpha$ is chosen in the range $(0.01, 10)$ because, even taking the full range of $\alpha$ values reported in Table~\ref{A_bounds}, no significant transverse spread is observed in the curves which look like one, thick curve. This result is consistent with the small $\alpha$-dependent corrections predicted by the expansion in eq.~(\ref{TreAexpan}). Note that we do not use in the plot the bounds exactly as specified in Table~\ref{A_bounds}, as the goal is to understand the general evolution of the quantities involved. Note how tightly constrained the temperature $T_{re}$ is.}
\label{A_Tre_Nk}
\end{center}
\end{figure*}

We can refine our calculation by removing the assumption $ V(\phi_k)\approx V_0$ in eq.~(\ref{V01}). Using eq.~(\ref{A1}), we can solve for $V_0$ in terms of $\phi_k$ and the amplitude of scalar perturbations $A_s$ at horizon crossing:
\begin{equation}
\label{V0}
V_0 = \frac{12 A_s \pi^2 \left(f^{\prime}(\phi_k) M_{Pl}\right)^2}{f(\phi_k)^3} M_{Pl}^4,
\end{equation}
where we have written again $V(\phi) = V_0 f(\phi)$. Here, $f^{\prime}(\phi_k)$ denotes the derivative of $f(\phi)$ with respect to $\phi$, evaluated at $\phi = \phi_k$. A general expression for the reheating temperature is then given by
\begin{equation}
\label{TreGeneral}
T_{re} \approx \left(\frac{135A_s^3 \pi^2 f^{\prime}(\phi_k)^6 \beta^3 M_{Pl}^{12}}{32 g_{re} f(\phi_k)^9}\right)^{1/4} M_{Pl}.
\end{equation}
For the $\alpha$-attractor model, where $\beta = \frac{1}{3\alpha M_{Pl}^2}$, the functions $f(\phi_k)$ and $f^{\prime}(\phi_k)$ are given by 
\begin{equation}
\label{efesattractor}
f(\phi_k) = \text{tanh}^2\left(\frac{\phi_k}{\sqrt{6\alpha} M_{Pl}}\right), \quad f^{\prime}(\phi_k) = \frac{1}{M_{Pl}} \sqrt{\frac{2}{3\alpha}} \, \text{sech}^2\left(\frac{\phi_k}{\sqrt{6\alpha} M_{Pl}}\right) \, \text{tanh}\left(\frac{\phi_k}{\sqrt{6\alpha} M_{Pl}}\right).
\end{equation}
From eq.~(\ref{TreGeneral}), we find that, for the $\alpha$-attractor model defined by eq.~(\ref{A_pot})
\begin{equation}
\label{TreA1}
T_{re} \approx \left(\frac{5 A_s^3 \pi^2}{108 g_{re} \alpha^6}\right)^{1/4} \text{csch}^3\left(\frac{\phi_k}{\sqrt{6\alpha} M_{Pl}}\right) M_{Pl}.
\end{equation}
To study the  large $N_k$ behavior of $T_{re}$ we can determine $\phi_k$ by inverting the expression for the number of $e$-folds during inflation as in eq.~(\ref{fik}). This yields
\begin{equation}
\label{TreAexpan}
T_{re} \approx \left(\frac{135 A_s^3 \pi^2}{256 g_{re}}\right)^{1/4} \frac{M_{Pl}}{N_k^{3/2}} \left(1 + \frac{9\alpha}{8N_k}\left(1 - \sqrt{\frac{4+3\alpha}{3\alpha}}\right) + \cdots \right).
\end{equation}
Similarly, for the $\alpha$-Starobinsky model, where we also arrive at eq.~(\ref{TreA1}), the expansion of $T_{re}$ is
\begin{equation}
\label{TreSexpan}
T_{re} \approx \sqrt{8}\left(\frac{135 A_s^3 \pi^2}{256 g_{re}}\right)^{1/4} \frac{M_{Pl}}{N_k^{3/2}} \left(1 - \frac{9\alpha}{8N_k} \left(\ln \frac{4N_k}{3\alpha} + 1 + \ln\frac{2}{\sqrt{3\alpha}} - \ln\left(1 + \frac{2}{\sqrt{3\alpha}}\right) + \cdots \right)\right).
\end{equation}
where, in this case, $\phi_k$ is given by eq.~(\ref{fikS}), obtained by inverting eq.~(\ref{NkS}). It is evident that the leading term in the expansion (\ref{TreAexpan}) corresponds to eq.~(\ref{TreA}), while the second, $\alpha$-dependent term can be neglected. On the other hand, the expansion (\ref{TreSexpan}) is well approximated by the leading term (\ref{TreS}) for small values of $\alpha$ only, showing a broad range of $T_{re}$ values as $\alpha$ varies. This can be observed in the figures below. Both expansions, at leading order in the large-$N_k$ limit, are independent of $\alpha$. Figures~\ref{A_Tre_Nk} and \ref{S_Tre_Nk} show the behavior of $T_{re}$ with respect to $N_k$ for the $\alpha$-attractor and the $\alpha$-Starobinsky models, respectively. The common $N_k$-dependence in both models is another example of the universality within this class of models.
\section{The \boldmath{$\alpha$}-attractor model}\label{Attractor}

In this section, we examine the inflationary model given by the potential
\begin{equation}
V(\phi) = V_0\,  \text{tanh}^2\left(\frac{\phi}{\sqrt{6\alpha}M_{Pl}}\right),
\label{A_pot}
\end{equation}
where $\alpha$ is a dimensionless parameter (for a sample of papers with emphasis on $\alpha$-attractors see \cite{Linder:2015qxa}-\cite{Rodrigues:2020fle} and in relation with reheating \cite{Martin:2014nya}-\cite{Drewes:2023bbs}). A notable feature of the potential~(\ref{A_pot}) is its quadratic form around its minimum. In this region, the potential can be expanded as
\begin{equation}
\frac{V(\phi)}{V_0} = \left(\frac{\phi}{\sqrt{6\alpha}M_{Pl}}\right)^2 - \frac{2}{3}\left(\frac{\phi}{\sqrt{6\alpha}M_{Pl}}\right)^4 + \dots.
\label{potorigins}
\end{equation}
Works which primarily focus on Bayesian analyses and numerical determinations of bounds for various cosmological quantities are \cite{Cedeno:2019cgr}-\cite{Ballardini:2024ado}. In this paper, we build upon these studies by providing a detailed illustration of the evolution of observables and important cosmological quantities as functions of the EoS parameter, $\omega_{re}$, and the scalar spectral index, $n_s$.  This is achieved by numerically solving eq.~(\ref{Nk2}) for the inflaton field at horizon crossing, $\phi_k$, across a range of values for the parameter $\alpha$ and within the EoS interval $-1/3 < \omega_{re} < 1$. Additionally, we incorporate the expression for the reheating temperature, eq.(\ref{TreA1}), into eq.~(\ref{Nk2}).

The number of $e$-folds during inflation, $N_{k} = -\frac{1}{M_{Pl}^2}\int_{\phi_k}^{\phi_{end}}\frac{V}{V'}d\phi$, is given by
\begin{equation}
N_k = \frac{3}{2}\alpha\left(\cosh^2\left(\frac{\phi_k}{\sqrt{6\alpha}M_{Pl}}\right) - \cosh^2\left(\frac{\phi_e}{\sqrt{6\alpha}M_{Pl}}\right)\right).
\label{NkA1}
\end{equation}
By setting $\epsilon = 1$, the value of $\phi_{end}$ is found to be
\begin{equation}
\phi_{end} = \sqrt{\frac{3\alpha}{2}}M_{Pl}\sinh^{-1}\left(\frac{2}{\sqrt{3\alpha}}\right).
\label{fieA}
\end{equation}
Eq.~(\ref{NkA1}) can be solved for $\phi_k$, yielding 
\begin{equation}
\phi_k = \sqrt{\frac{3\alpha}{2}} M_{Pl}\cosh^{-1}\left(\sqrt{1 + \frac{4}{3\alpha}} + \frac{4 N_k}{3\alpha}\right).
\label{fik}
\end{equation}
Connections to inflationary models are made through cosmological observables, which to first order in the slow-roll (SR) approximation, are given by (see, e.g., \cite{Liddle:1994dx}, \cite{Lyth:1998xn})
\begin{eqnarray}
n_{s} &=&1+2\eta -6\epsilon ,  \label{Ins} \\
n_{sk}\equiv \frac{d n_s}{d \ln k} &=&16\epsilon \eta -24\epsilon ^{2}-2\xi_2, \label{Insk} \\
n_{t} &=&-2\epsilon, \label{Int} \\
n_{tk}\equiv \frac{d n_t}{d \ln k} &=&4\epsilon\left( \eta -2\epsilon\right), \label{Intk}
\end{eqnarray}
where $n_{sk}$ (often denoted by $\alpha$) is the running of the scalar spectral index, $n_t$ is the tensor spectral index, and $n_{tk}$ the running of the tensor index. All quantities are evaluated at horizon crossing for wavenumber $k$. The slow-roll parameters used above are defined as
\begin{equation}
\epsilon \equiv \frac{M_{Pl}^{2}}{2}\left( \frac{V^{\prime }}{V }\right) ^{2},\quad\quad
\eta \equiv M_{Pl}^{2}\frac{V^{\prime \prime }}{V}, \quad\quad
\xi_2 \equiv M_{Pl}^{4}\frac{V^{\prime }V^{\prime \prime \prime }}{V^{2}},
\label{Spa}
\end{equation}
where primes indicate derivatives of $V$ with respect to the inflaton field $\phi$. The observables from eqs.~(\ref{Ins})-(\ref{Intk}) can be computed, resulting in
\begin{eqnarray}
n_{s} &=&1-\frac{4}{3\alpha}\csch^2\left(\frac{\phi_k}{\sqrt{6\alpha}M_{Pl}}\right) ,  \label{OIns} \\
n_{sk}&=&-\frac{8}{9\alpha^2}\csch^4\left(\frac{\phi_k}{\sqrt{6\alpha}M_{Pl}}\right), \label{OInsk} \\
n_{t} &=&-\frac{8}{3\alpha}\csch^2\left(\sqrt{\frac{2}{3\alpha}}\frac{\phi_k}{M_{Pl}}\right), \label{OInt} \\
n_{tk}&=&-\frac{64}{9\alpha^2}\coth\left(\sqrt{\frac{2}{3\alpha}}\frac{\phi_k}{M_{Pl}}\right)\csch^3\left(\sqrt{\frac{2}{3\alpha}}\frac{\phi_k}{M_{Pl}}\right), \label{OIntk}
\end{eqnarray}
respectively. Substituting $\phi_k$ from eq.~(\ref{fik}) into eqs.~(\ref{OIns})-(\ref{OIntk}) and (\ref{TreA1}), and performing a large-$N_k$ expansion, we obtain, to leading order
\begin{equation}
\label{largeNk}
n_s=1-\frac{2}{N_k}, \quad n_{sk}=-\frac{2}{N_k^2},\quad n_t=-\frac{3\alpha}{2N_k^2}, \quad n_{tk}=-\frac{3\alpha}{N_k^3},  \quad T_{re}= \left(\frac{135 A_s^3 \pi^2}{256g_{re}}\right)^{1/4}\frac{M_{Pl} }{N_k^{3/2}} 
\end{equation}
Solving eq.~(\ref{OIns}) for $\phi_k$, we get
\begin{equation}
\label{fikns}
\phi_k=\sqrt{6\alpha}M_{Pl}\sinh^{-1}\left(\frac{2}{\sqrt{3\alpha \delta_{n_s}}}\right),
\end{equation}
where $\delta_{n_s} \equiv 1 - n_s$. Substituting $\phi_k$ from eq.~(\ref{fikns}) into $r=-8n_t=\frac{64}{3\alpha}\csch^2\left(\sqrt{\frac{2}{3\alpha}}\frac{\phi_k}{M_{Pl}}\right)$ and solving for $\alpha$ in terms of the observables $n_s$ and $r$, we obtain
\begin{equation}
\alpha = \frac{4r}{3\delta_{n_s}(4\delta_{n_s} - r)}.
\label{alfaA}
\end{equation}
Thus, the ranges of the observables $n_s$ and $r$ can be used to determine the range of $\alpha$. Eliminating $\alpha$ with (\ref{alfaA}) we can also write
\begin{eqnarray}
n_{sk}&=&-\frac{\delta_{n_s}^2}{2}, \label{2Insk} \\
n_{tk}&=&-\frac{1}{64}r\left(8\delta_{n_s}-r\right), \label{2OIntk}
\end{eqnarray}
which are consistency relations of the model. Moreover, the number of $e$-folds during inflation can be expressed as  \cite{German:2020cbw}
\begin{equation}
N_k = \frac{8\delta_{n_s}-r-\sqrt{r^2+r\delta_{n_s}(4\delta_{n_s}-r)}}{\delta_{n_s}(4\delta_{n_s} - r)},
\label{NkA2}
\end{equation}
and the reheating temperature is given by
\begin{equation}
\label{TreA2}
T_{re}\approx \left(\frac{135A_s^3\pi^2}{256g_{re}}\right)^{1/4} \frac{M_{Pl}}{\sqrt{8}}\,\delta_{n_s}^{3/2},
\end{equation}
Following the approach in \cite{German:2023yer}, but extending the range of $\omega_{re}$ and, more importantly, incorporating the dynamical $T_{re}$ of eq.~(\ref{TreA1}), we initially adopt the bounds from Table 3 of Ref.~\cite{Akrami:2018odb}. These bounds correspond to the cosmological model $\Lambda$CDM$+r+dn_s/d\ln k$, based on the Planck TT, TE, EE + lowE + lensing + BK15 + BAO dataset. These constraints provide limits on the parameters and observables relevant to the specified cosmological model and dataset
\begin{equation}
n_s = 0.9658 \pm 0.0040 \quad (68\%\,\, \text{C.L.}),
\label{boundsns} 
\end{equation}
\begin{equation}
r < 0.068 \quad (95\%\,\, \text{C.L.}).
\label{boundcr} 
\end{equation}
\begin{figure*}[h!]
\begin{center}
\includegraphics[width=4.5 in]{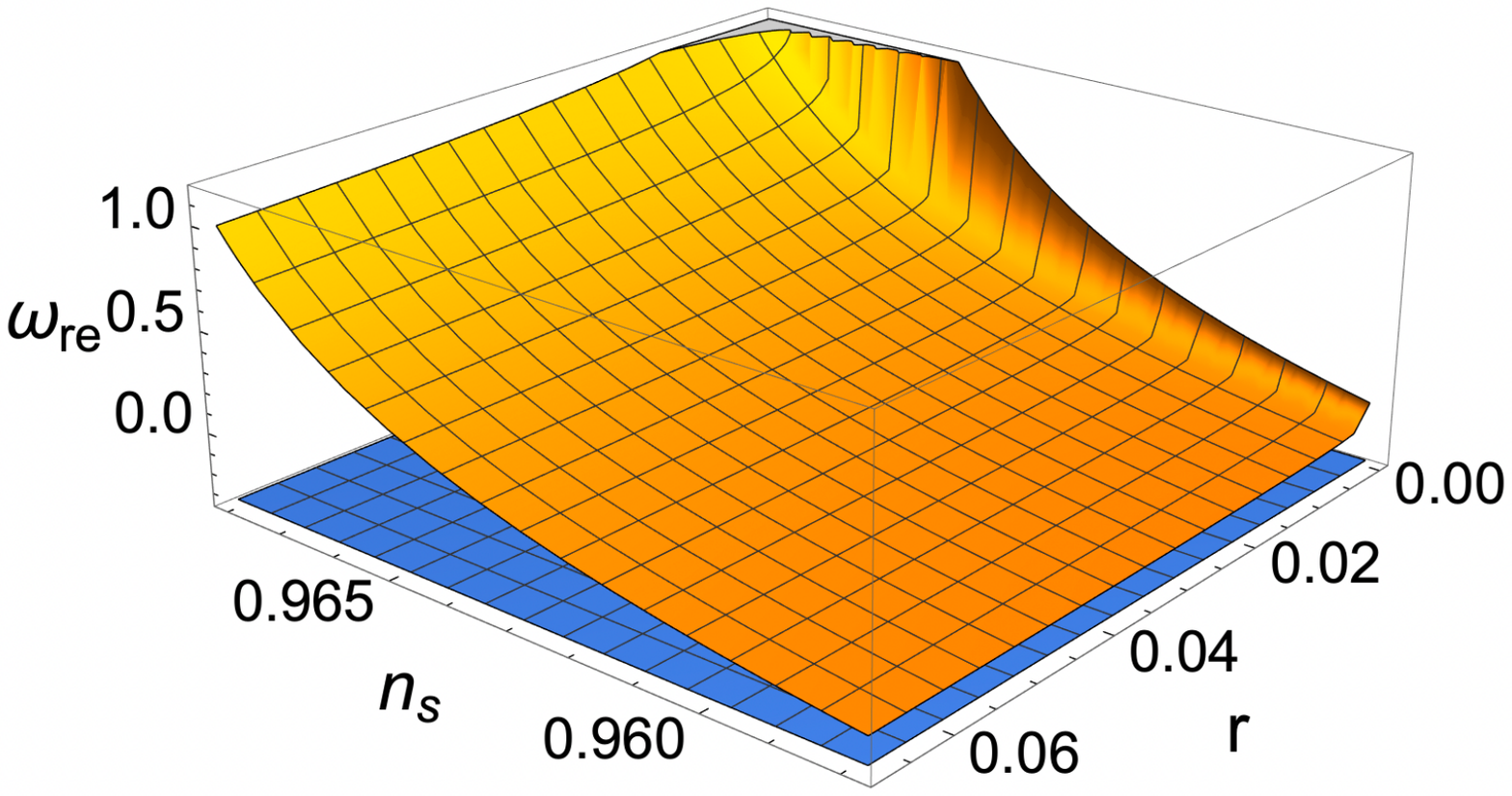}
\caption{The figure shows the EoS parameter $\omega_{re}$, as given by eq.~(\ref{omega1}), as a function of $n_s$ and $r$ for the $\alpha$-attractor class of models defined by eq.~(\ref{A_pot}). The range $-\frac{1}{3} < \omega_{re} < 1$, in combination with $0.9578 < n_s < 0.9738$ and $0 < r < 0.068$, results in constrained ranges for both the EoS and the observables, as discussed in the text and summarized in Table~\ref{A_bounds}.}
\label{A_wre_ns_r}
\end{center}
\end{figure*}
\begin{table*}[htbp!]
 \begin{center}
{\begin{tabular}{cccc}
\small
Parameter & Value & Parameter & Value  \\
\hline \hline
{$g_{re} $} & $106.75$ & {$k_p $} & $0.05Mpc^{-1}$ \\
{$g_{s,re} $} & $106.75$ & {$T_0 $}   & $2.7255\,K$  \\
{$A_s $}  & $2.1\times 10^{-9}$ & {$a_{eq} $}& $2.94\times 10^{-4}$ \\
 \hline \hline
Observable & range & others & range  \\
\hline \hline
{$\omega_{re}$} & $(-0.20,1)$ & $\alpha$ & $(5\times 10^{-12}, 47.4)$  \\
{$n_{s}$} & $(0.9578, 0.9680)$ & $N_{k}$ & $(46.9, 62.1)$  \\
{$r$} & $(2.67\times 10^{-14}, 0.068)$ & $N_{re}$ & $(24.8, 10.1)$ \\
{$n_{sk}$} & $(-0.00089, -0.00051)$ & $N_{rd}$ & $(42.9, 42.4)$\\
{$n_{tk}$} & $(-0.00029, -0.00020)$ & $T_{re}/\mathrm{GeV}$ & $(1.1\times 10^{9}, 7.2\times 10^{8})$  \\
\hline\hline
\end{tabular}}
\caption{For the $\alpha$-attractor model defined by eq.~(\ref{A_pot}), the table above provides the parameter values used in the calculations, while the table below lists the range of values for the observables: the scalar spectral index $n_s$, the tensor-to-scalar ratio $r$, the running of the scalar spectral index $n_{sk}$, the running of the tensor spectral index $n_{tk}$. The parameter $\alpha$, the number of $e$-folds during inflation $N_k$, reheating $N_{re}$, and radiation $N_{rd}$, as well as the reheating temperature $T_{\text{re}}$, which is tightly constrained.
The lower bound on $n_s$ ($n_s = 0.9578$, at $2\sigma$) implies a lower bound on $\omega_{\text{re}}$ ($\omega_{\text{re}} = -0.20$), while the upper bound on $\omega_{\text{re}}$ ($\omega_{\text{re}} < 1$) requires $n_s < 0.9680$ and $r > 2.67 \times 10^{-14}$. Note that most of the figures shown below extend these ranges for illustrative purposes.
A recently reported most probable value for vanilla inflation, $n_{sk} = -6.3 \times 10^{-4}$ \cite{Martin:2024nlo}, falls comfortably within the range above.}
\label{A_bounds}
\end{center}
\end{table*}
\begin{figure*}[h!]
\begin{center}
$\begin{array}{ccc}
\includegraphics[width=3.in]{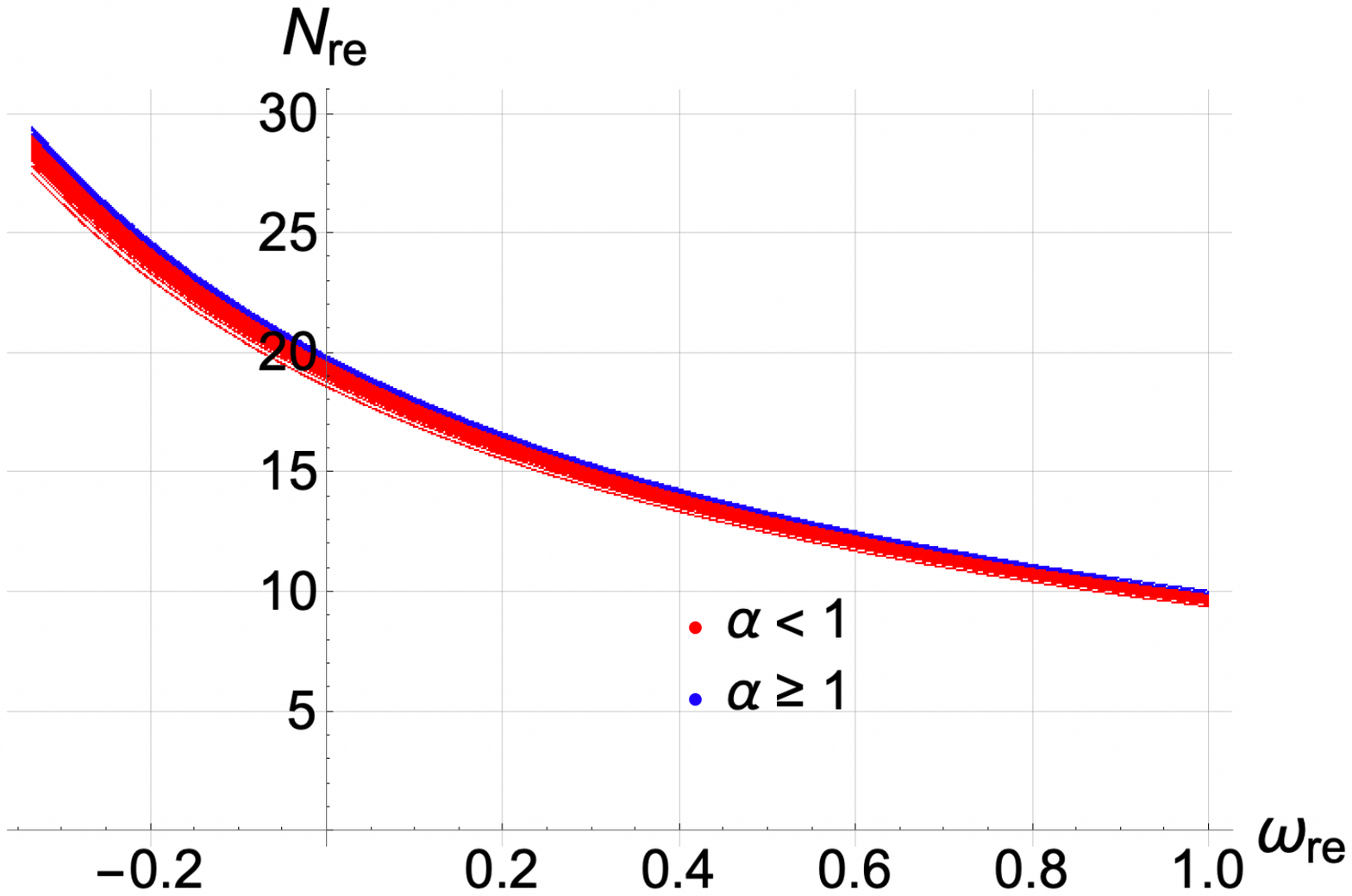}&
\includegraphics[width=3.in]{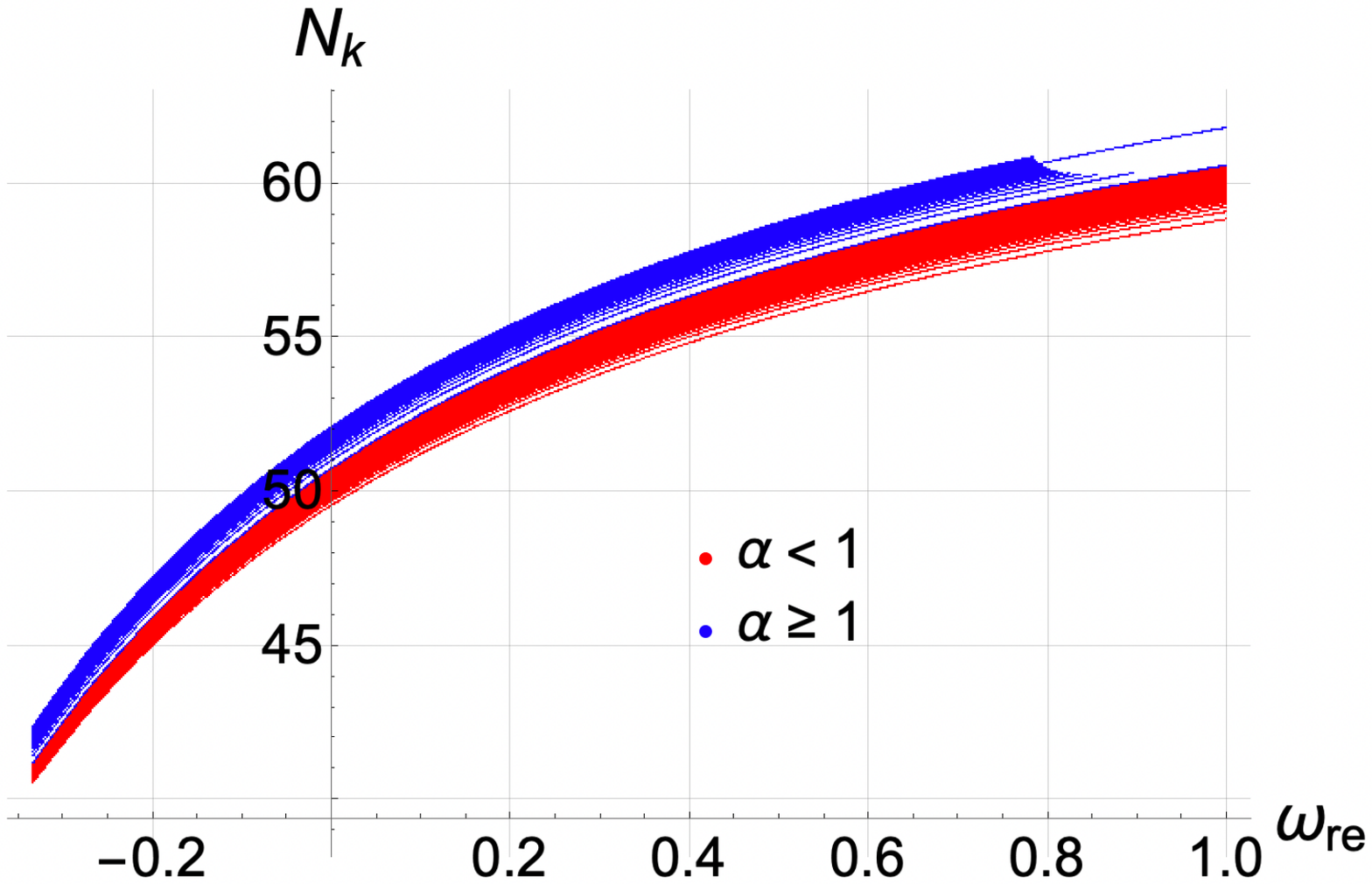}\\
\end{array}$
\caption{The plots show the number of $e$-folds during reheating, $N_{re}$, and during inflation, $N_k$, as functions of the EoS parameter $\omega_{re}$ for the $\alpha$-attractor model of eq.~(\ref{A_pot}). Shaded regions differentiate solutions with $\alpha < 1$ from those with $\alpha > 1$, based on the model parameter $\alpha$. Other details are the same as in Fig.~\ref{A_Tre_Nk}.}
\label{A_Nk_Nre_wre}
\end{center}
\end{figure*}
We extend the range of $n_s$ up to $2\sigma$. To incorporate reheating constraints into inflation, we plot $\omega_{re}$ from eq.~(\ref{omega1}) as a function of $n_s$ and $r$. This is achieved by substituting eqs.~(\ref{roend})-(\ref{Nkeq2}) and (\ref{TreA1}) into (\ref{omega1}), along with eqs.~(\ref{NkA1}), (\ref{fieA}),  (\ref{fikns}), and (\ref{alfaA}). The resulting surface for $\omega_{re}$ is shown in Fig.~\ref{A_wre_ns_r}. From this, we infer that the range $-\frac{1}{3}<\omega_{re}<1$, combined with $0.9578<n_s<0.9738$ and $0< r <0.068$, leads to constrained ranges for both the EoS and the observables, as summarized and detailed in Table~\ref{A_bounds}.
Up to this point, we have discussed the model defined by eq.~(\ref{A_pot}) without addressing its fundamental origin. However, in gravitational theories, including supersymmetric models and frameworks related to $\alpha$-attractors, the gravitino problem is a well-known issue~\cite{Ellis:1982yb}. To avoid this problem, the reheating temperature must be limited to values of the order of  $10^9$ GeV to prevent the overproduction of gravitinos and other relic particles, which could jeopardize the success of Big Bang nucleosynthesis~\cite{Kawasaki:2006gs, Kawasaki:2006hm, Kohri:2005wn}. It is therefore reassuring that the upper bound on $T_{re}$ in Table~\ref{A_bounds} satisfies this requirement.

\begin{figure*}[h]
\begin{center}
$\begin{array}{ccc}
\includegraphics[width=3.in]{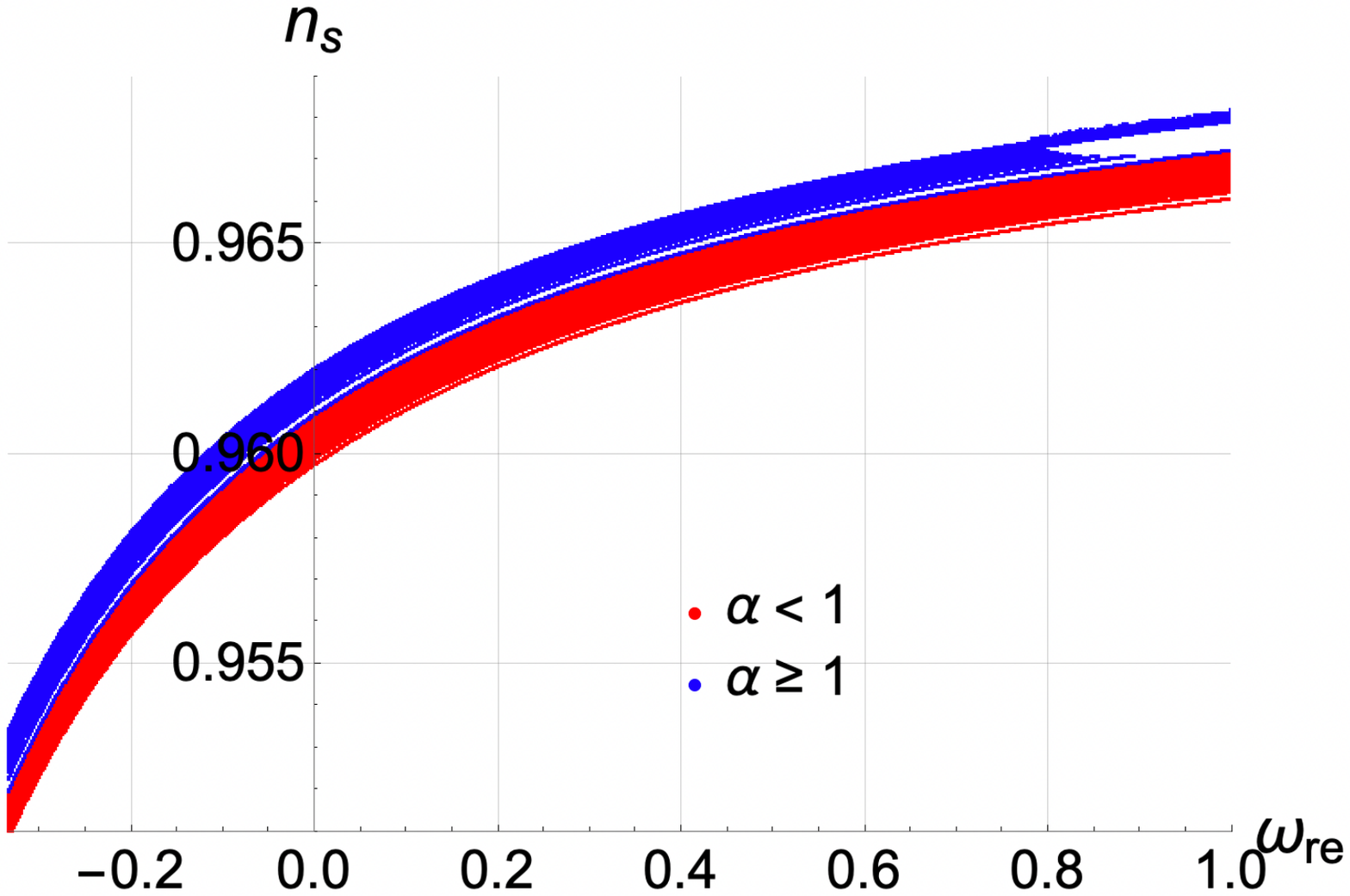}&
\includegraphics[width=3.in]{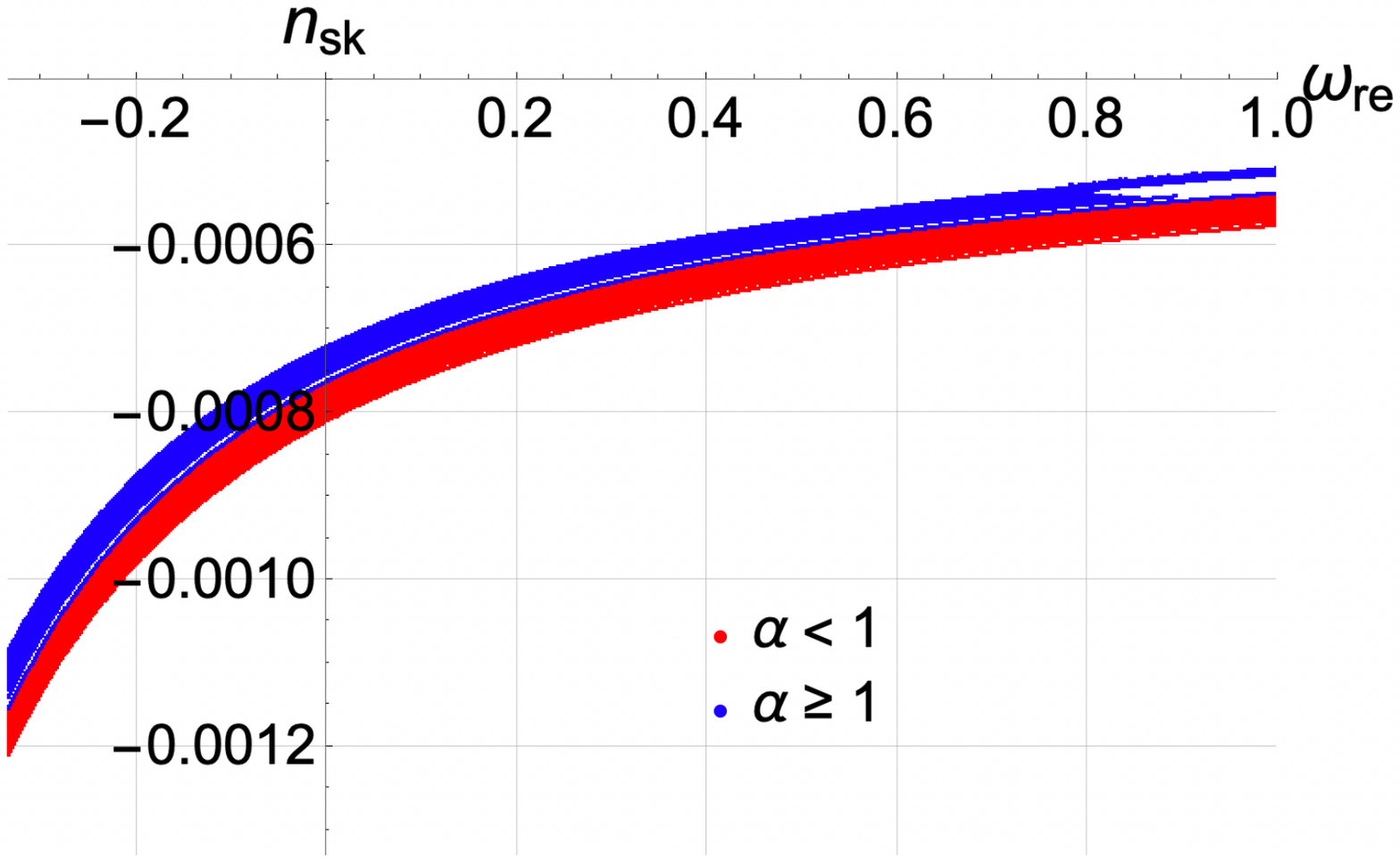}\\
\end{array}$
\caption{The plots show the scalar spectral index, $n_s$, and its running, denoted here by $n_{sk}$, as functions of the EoS parameter $\omega_{re}$ for the $\alpha$-attractor model defined by eq.~(\ref{A_pot}). Shaded regions distinguish solutions with $\alpha < 1$ from those with $\alpha > 1$, according to the model parameter $\alpha$. Other details are the same as in Fig.~\ref{A_Tre_Nk}.}
\label{A_ns_nsk_wre}
\end{center}
\end{figure*}

\begin{figure*}[h]
\begin{center}
\includegraphics[width=4.in]{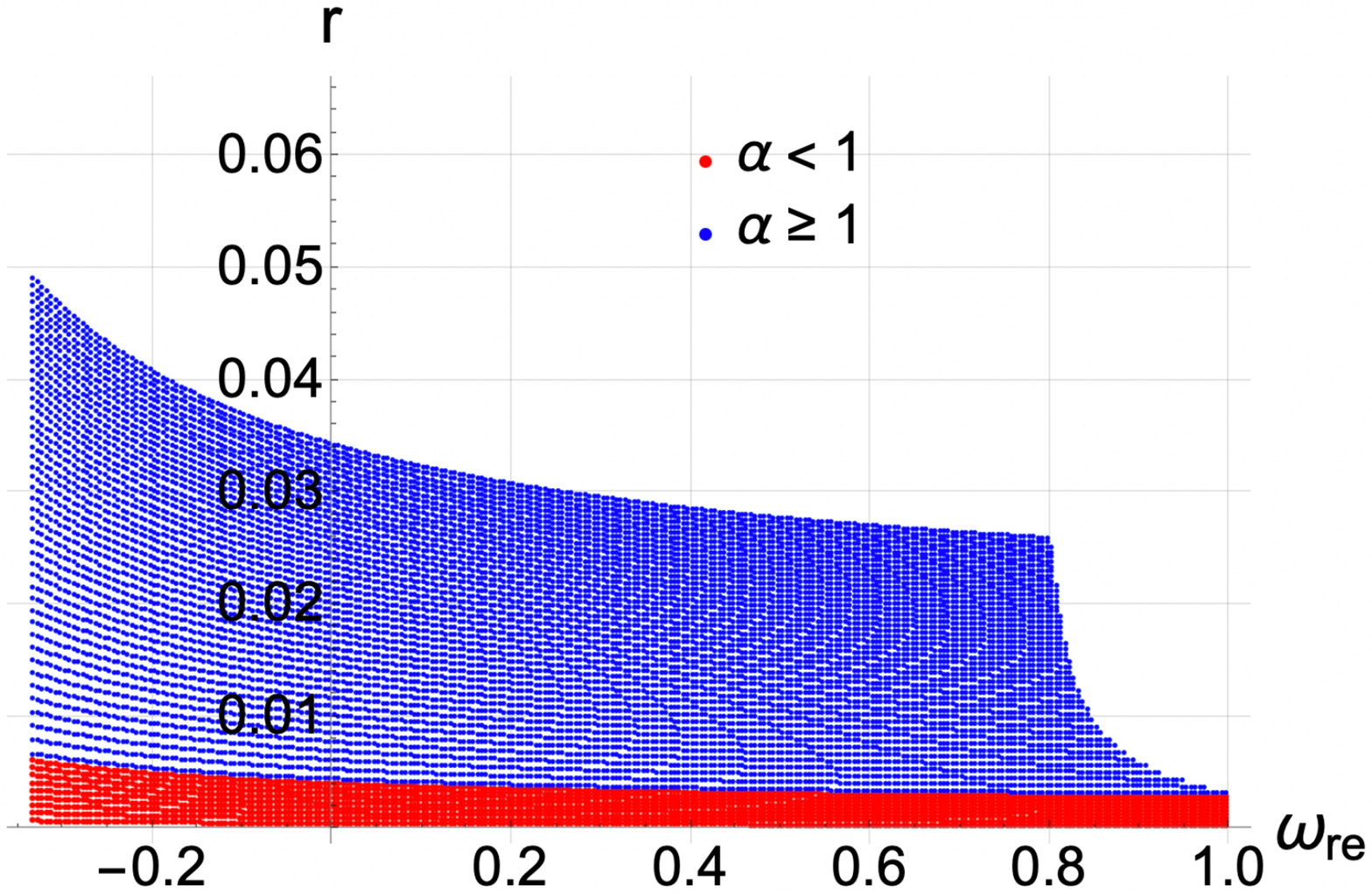}
\caption{The figure shows the tensor-to-scalar ratio, $r$, as a function of the EoS parameter $\omega_{re}$ for the $\alpha$-attractor model defined by eq.~(\ref{A_pot}). Shaded regions differentiate solutions with $\alpha < 1$ from those with $\alpha > 1$, based on the model parameter $\alpha$. Other details are the same as in Fig.~\ref{A_Tre_Nk}.}
\label{A_r_wre}
\end{center}
\end{figure*}

\begin{figure*}[p]
\begin{center}
$\begin{array}{ccc}
\includegraphics[width=3.in]{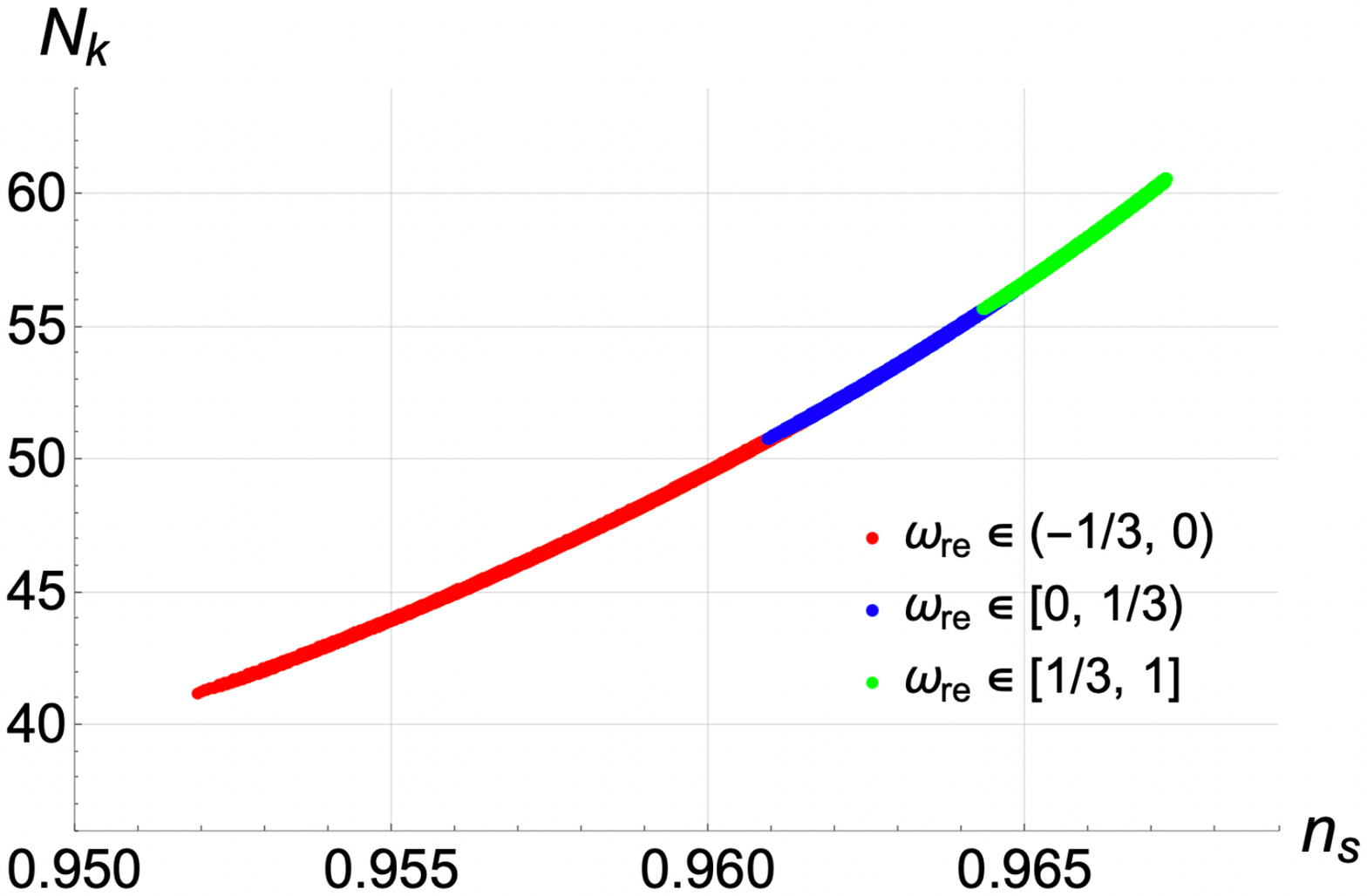}&
\includegraphics[width=3.in]{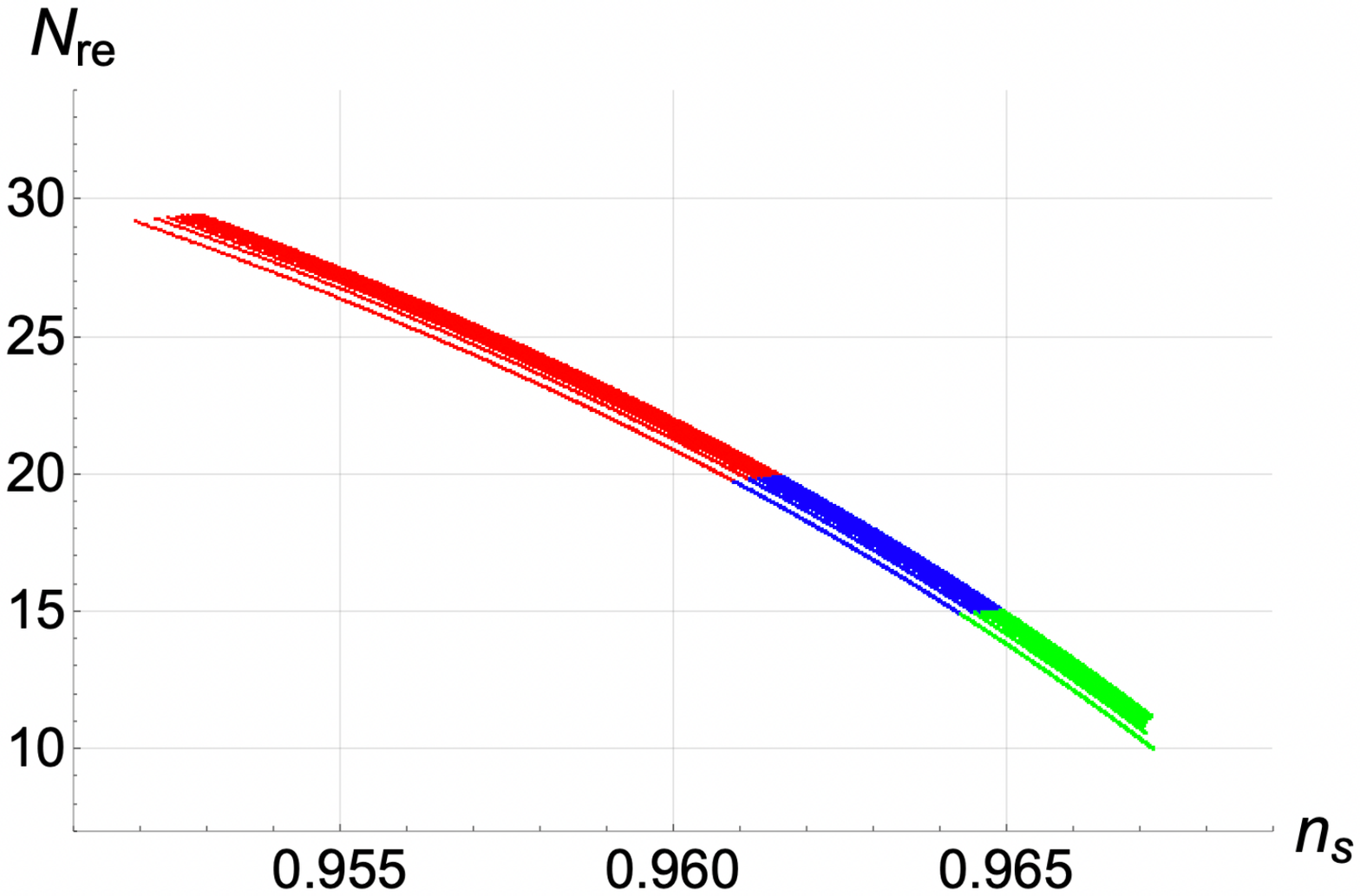}\\
\includegraphics[width=3.in]{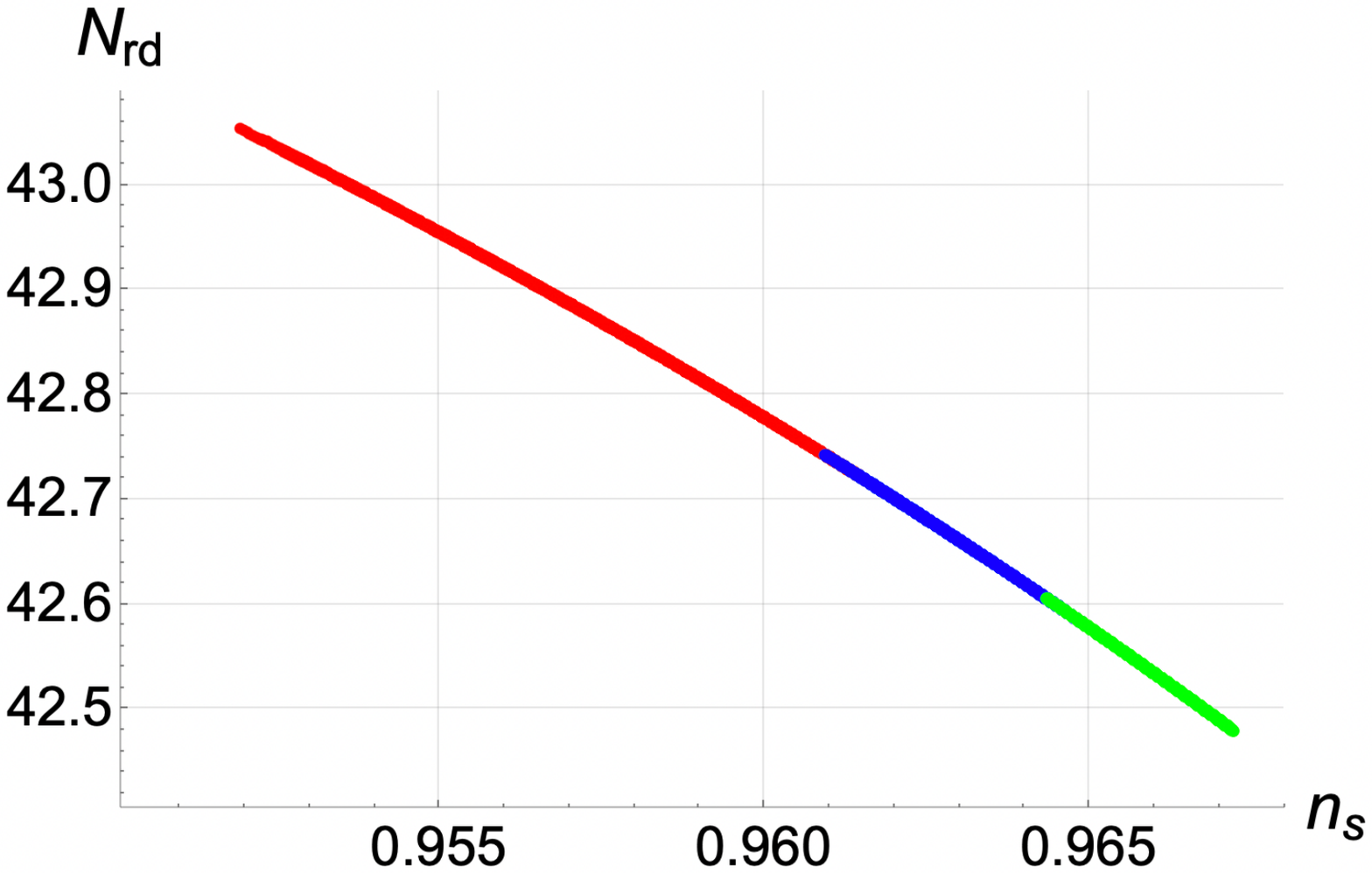}&
\includegraphics[width=3.in]{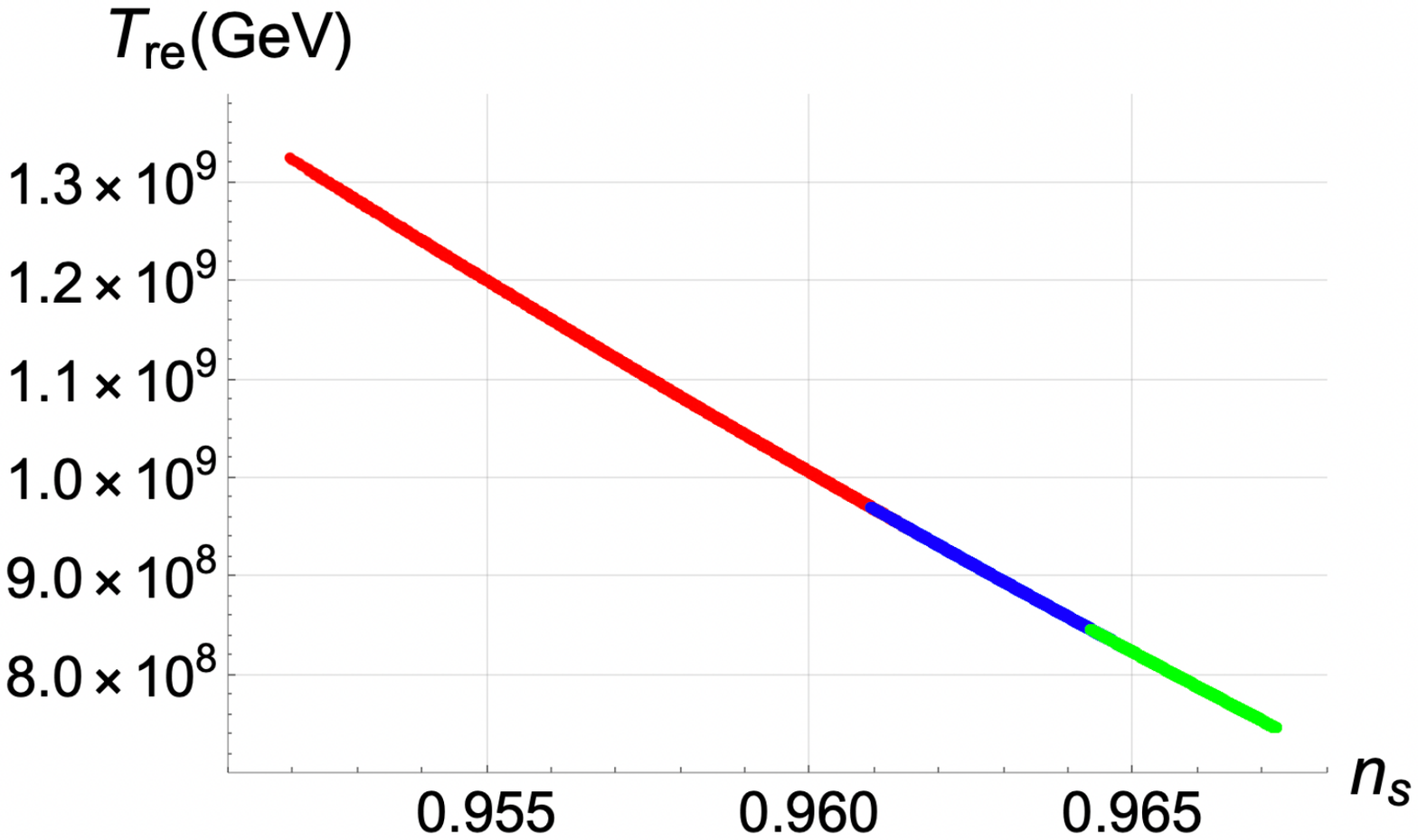} \\
\end{array}$
\caption{The plots show the number of $e$-folds during inflation, $N_k$, reheating, $N_{re}$, and during radiation, $N_{rd}$, as well as the reheating temperature at the end of reheating, $T_{re}$, as functions of the scalar spectral index, $n_s$, for the $\alpha$-attractor model defined by eq.~(\ref{A_pot}). Details are the same as in Fig.~\ref{A_Tre_Nk}.}
\label{A_efolds_ns}
\end{center}
\end{figure*}

\begin{figure*}[p]
\begin{center}
$\begin{array}{ccc}
\includegraphics[width=3.in]{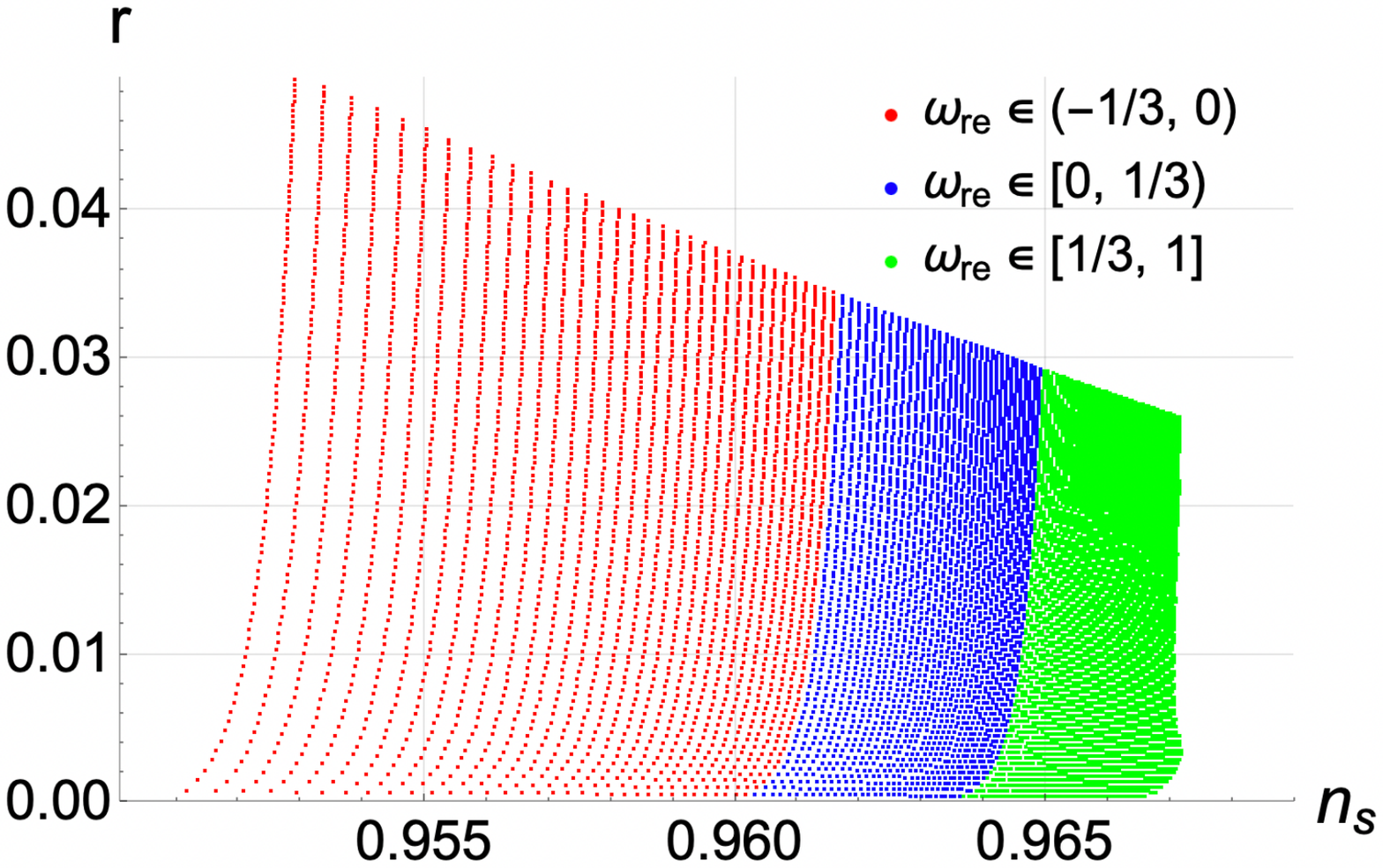}&
\includegraphics[width=3.in]{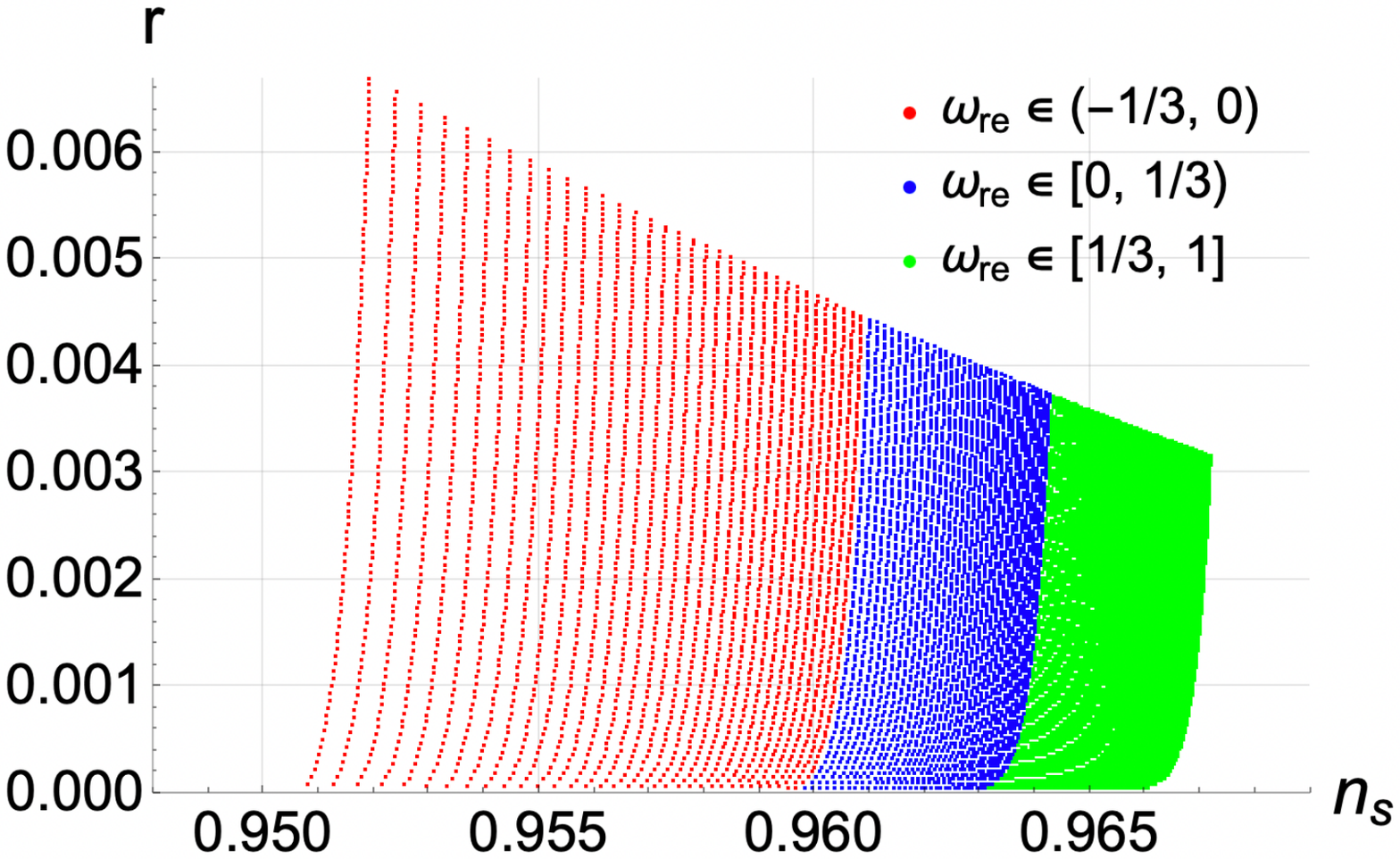}\\
\end{array}$
\caption{Plots of the tensor-to-scalar ratio, $r$, as a function of the scalar spectral index, $n_s$, for the $\alpha$-attractor model defined by eq.~(\ref{A_pot}). The figure on the right-hand side corresponds to $\alpha \leq 1$ values of the model parameter. The values of $r$ increase as $\alpha$ increases in value. Other details are the same as in Fig.~\ref{A_Tre_Nk}.}
\label{A_cosas_ns}
\end{center}
\end{figure*}

\begin{figure*}[h!]
\begin{center}
\includegraphics[width=4.in]{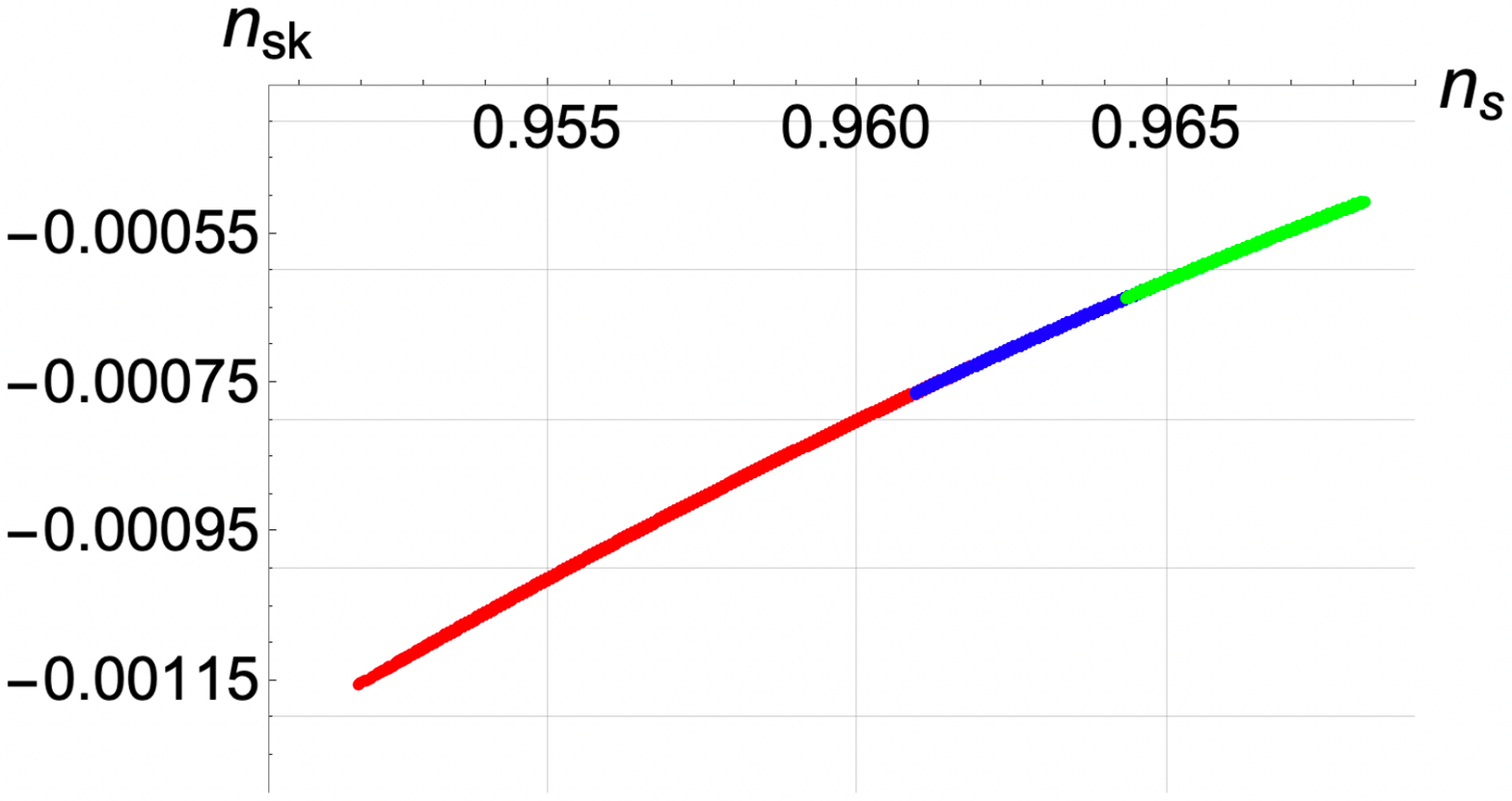}
\caption{Plot of the running of the scalar spectral index, denoted here by $n_{sk}$, as a function of $n_s$. for the $\alpha$-attractor model defined by eq.~(\ref{A_pot}). Details are the same as in Fig.~\ref{A_Tre_Nk}.}
\label{A_nsk_ns}
\end{center}
\end{figure*}

Table~\ref{A_bounds} presents the results of these calculations for the $\alpha$-attractor model from eq.~(\ref{A_pot}), while Figures \ref{A_Tre_Nk} to \ref{A_nsk_ns} illustrate the evolution of various cosmological quantities, as described in their respective captions, with general details provided in Figure~\ref{A_Tre_Nk}.
\section{\boldmath The $\alpha$-Starobinsky model}\label{Staro}
In this section, we examine the $\alpha$-Starobinsky model, which extends the original Starobinsky model by introducing a new parameter, $\alpha$. Our method follows the same procedure as outlined in the previous section. The potential for the $\alpha$-Starobinsky model is given by~\cite{Ellis:2013nxa,Kallosh:2013yoa,Ellis:2019bmm}
\begin{equation}
\label{VS}
 V(\phi) = V_{0}\left(1 - e^{-\sqrt{\frac{2}{3\alpha}}\frac{\phi}{M_{Pl}}}\right)^{2}.
\end{equation}
This generalized form is inspired by supergravity. In supergravity, the dynamics are determined by both the K{\"a}hler potential and the superpotential. For minimal supergravity models, the K{\"a}hler potential is given by \( K = \phi^{i}\phi_{i}^{*} \), where \(\phi_{i}\) are the chiral scalar fields in the theory. For simplicity, we set \( M_{Pl} = 1 \). Using this K{\"a}hler potential, the effective scalar field potential can be derived as
\begin{equation}
\label{Superpotential}
 V = e^{\phi^{i}\phi_{i}^{*}}\left(|W_{i} + \phi_{i}^{*}W|^{2} - 3|W|^{2}\right),
\end{equation}
where \( W(\phi) \) is the superpotential, a holomorphic function of the fields \(\phi_i\). The negative term in this potential means it is not necessarily positive, and in general, it lacks the flat region required for inflation. To resolve this issue, no-scale supergravity models were introduced~\cite{Cremmer:1983bf,Lahanas:1986uc}, which use the K{\"a}hler potential
$K = -3\ln \left(T + T^* - \frac{|\phi_{i}|^{2}}{3}\right)$,
where \(T\) is the volume modulus. When the modulus \(T\) is stabilized, the Starobinsky potential can be recovered. The Wess-Zumino superpotential~\cite{Ellis:2013xoa}
$W(\phi) = \mu\left(\frac{1}{2}\phi^2 - \frac{1}{3\sqrt{3}}\phi^3\right)$,
is introduced alongside other possible versions~\cite{Ellis:2013nxa,Ellis:2014gxa,Ellis:2015xna,Ellis:2020lnc}. In the $\alpha$-Starobinsky model, the K{\"a}hler potential is generalized to
\begin{equation}
\label{KP}
K = -3\alpha \ln \left(T + T^* - \frac{|\phi_{i}|^{2}}{3}\right),
\end{equation}
with a slightly more complex superpotential~\cite{Ellis:2019bmm}.

Given the inflaton's value at horizon exit, we can calculate the number of $e$-folds between horizon crossing and the end of inflation, $N_k$, using the slow-roll approximation
\begin{equation}
\label{NkS}
 N_{k}=-\frac{1}{M_{Pl}^{2}}\int _{\phi_{k}}^{\phi_{e}} \frac{V}{V'} d\phi = \frac{3\alpha}{4}\left(e^{\sqrt{\frac{2}{3\alpha}}\frac{\phi_{k}}{M_{Pl}}}-e^{\sqrt{\frac{2}{3\alpha}}\frac{\phi_{e}}{M_{Pl}}}+\sqrt{\frac{2}{3\alpha}}\left(\frac{\phi_{e}-\phi_{k}}{M_{Pl}}\right)\right),
\end{equation}
where $\phi_{e}$ is the value of the field at the end of inflation. For simplicity in calculations, we approximate this by setting $\epsilon \simeq 1$
\begin{equation}
\label{fieS}
\phi_{e}=\sqrt{\frac{3\alpha}{2}}M_{Pl}\ln \left(1+\frac{2}{\sqrt{3\alpha}}\right).
\end{equation}
The value of $\phi_k$ can be expressed as
\begin{equation}
\frac{\phi_k}{M_{Pl}} = \sqrt{\frac{3\alpha}{2}}\left(-\frac{4}{3\alpha}N_k - e^{\sqrt{\frac{2}{3\alpha}}\frac{\phi_e}{M_{Pl}}} + \sqrt{\frac{2}{3\alpha}}\frac{\phi_e}{M_{Pl}}\right) - \text{ProductLog}\left[-1, -e^{-\frac{4}{3\alpha}N_k - e^{\sqrt{\frac{2}{3\alpha}}\frac{\phi_e}{M_{Pl}}} + \sqrt{\frac{2}{3\alpha}}\frac{\phi_e}{M_{Pl}}}\right].
\label{fikS}
\end{equation}
Here, $\text{ProductLog}[-1,x]$ refers to the Lambert $W_{-1}$ function, as implemented in Mathematica, which corresponds to the secondary branch of the function. This expression is obtained by inverting eq.~(\ref{NkS}).

\begin{figure*}[h!]
\begin{center}
\includegraphics[width=4.5 in]{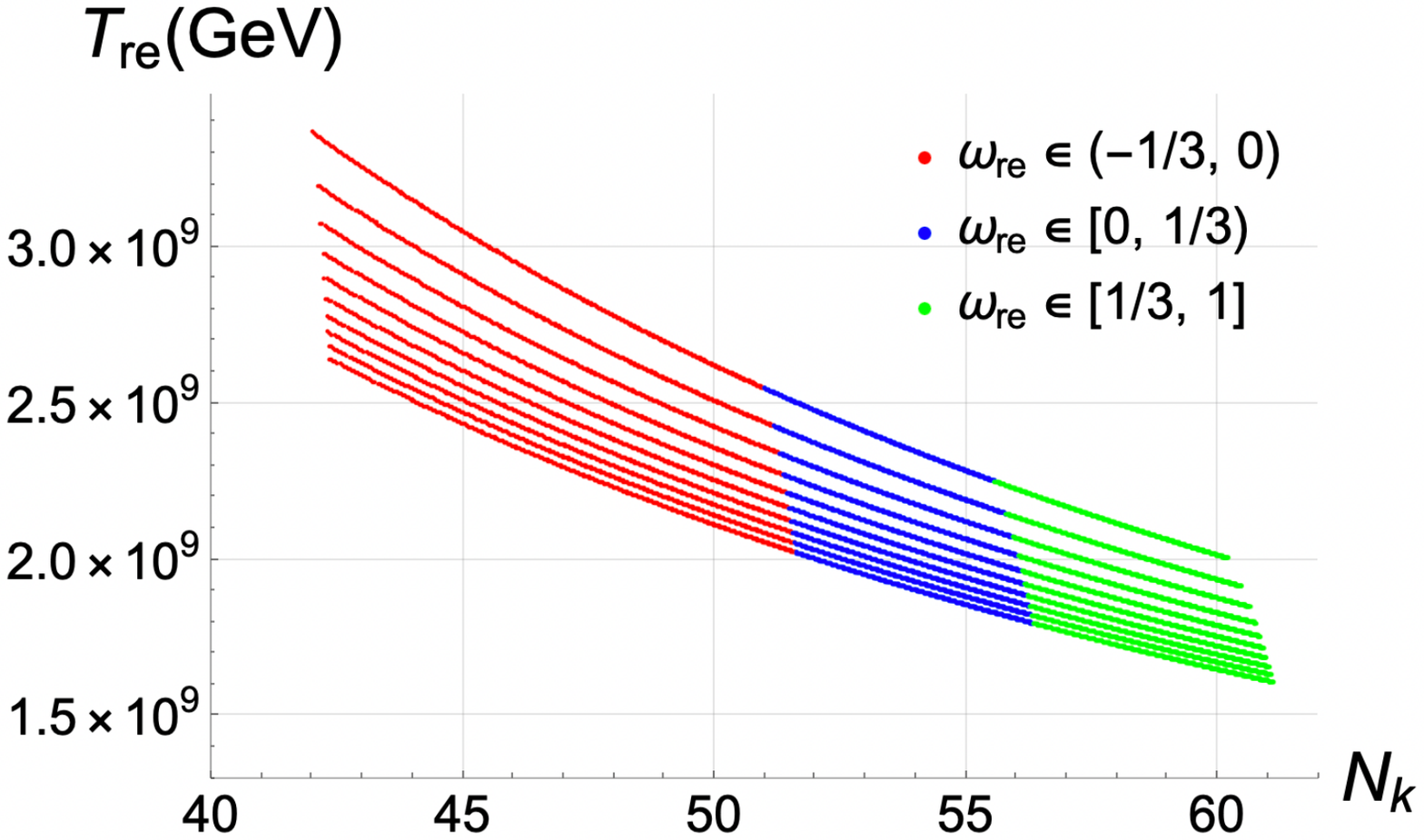}
\caption{The figure  is the counterpart for the $\alpha$-Starobinsky model of the Fig.~\ref{A_Tre_Nk} of eq.(\ref{TreA1}) supplemented with eq.~(\ref{fik}). The figure shows the number of $e$-folds during inflation, $N_k$, from eq.~(\ref{NkS}) plotted against the reheating temperature, $T_{re}$, from eq.~(\ref{TreA1}) for the $\alpha$-Starobinsky model defined by eq.~(\ref{VS}). The EoS interval $-1/3 < \omega_{re} < 1$ is divided into three regions, as specified. The parameter $\alpha$ is chosen in the range $(0.01, 10)$ with $T_{re}$ for each curve decreasing as $\alpha$ increases in value. We can see significant spread  of the curves across the $\alpha$ values listed in Table~\ref{S_bounds}. This is contrary to Fig.~\ref{A_Tre_Nk} where no spreading is observed. This result is consistent with the $\alpha$-dependent corrections predicted by the expansion in eq.~(\ref{TreSexpan}) where the correction to the leading term is small only for small $\alpha$. 
Note that we do not use the bounds from Table~\ref{S_bounds} in this plot, as the goal is to understand the general evolution of the quantities involved. The plot is generated by first solving the full eq.~(\ref{Nk2}) for $\phi_k$, with the plotted quantities derived from their respective defining equations. }
\label{S_Tre_Nk}
\end{center}
\end{figure*}

\begin{figure*}[p]
\begin{center}
$\begin{array}{ccc}
\includegraphics[width=3.in]{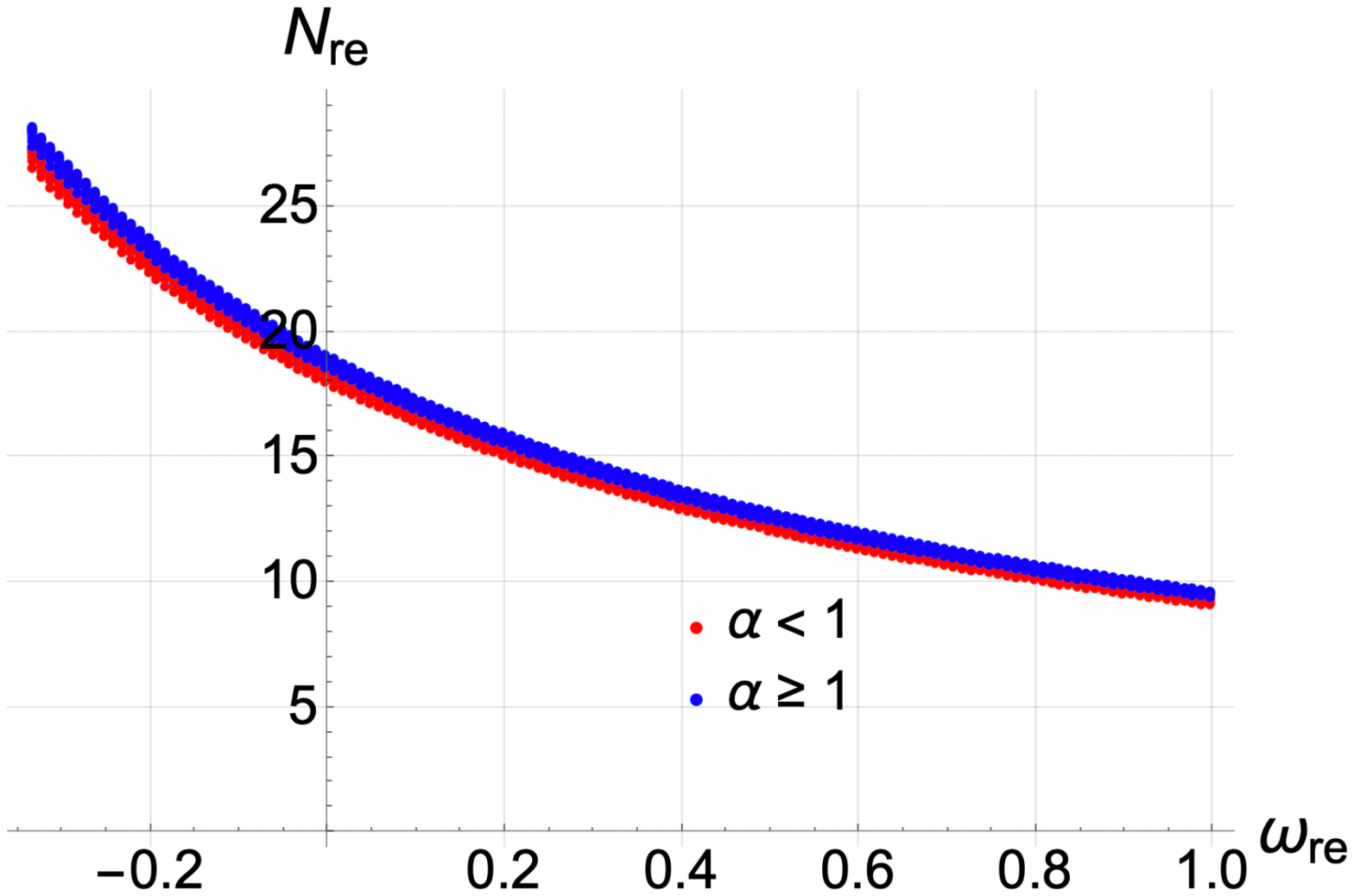}&
\includegraphics[width=3.in]{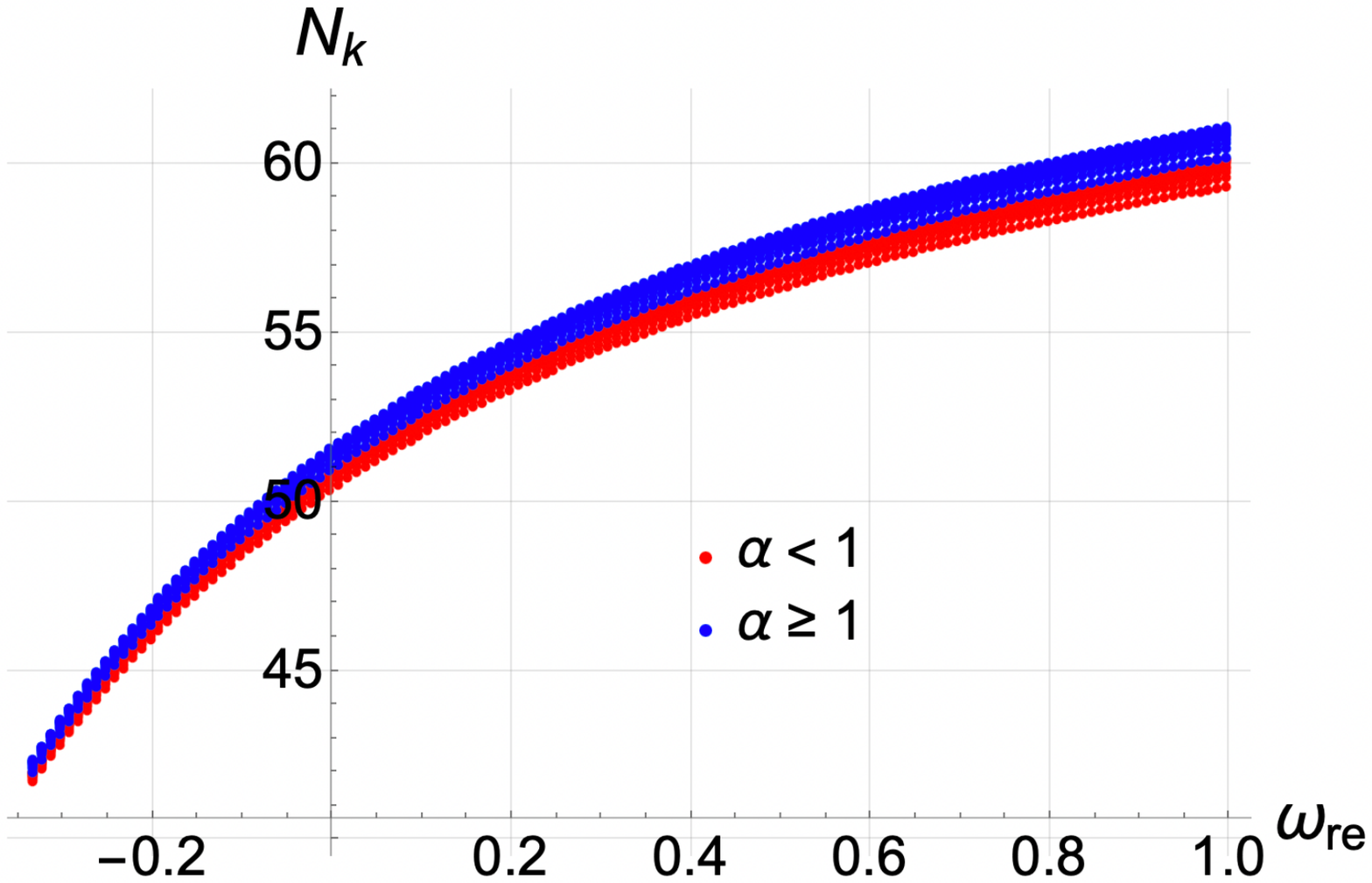}\\
\end{array}$
\caption{The plots show the number of $e$-folds during reheating, $N_{re}$, and during inflation, $N_k$, as functions of the EoS parameter $\omega_{re}$ for the $\alpha$-Starobinsky model of eq.~(\ref{VS}). Shaded regions differentiate solutions with $\alpha < 1$ from those with $\alpha > 1$, based on the model parameter $\alpha$. Other details are the same as in Fig.~\ref{S_Tre_Nk}.}
\label{S_Nk_Nre_wre}
\end{center}
\end{figure*}

\begin{figure*}[p]
\begin{center}
$\begin{array}{ccc}
\includegraphics[width=3.in]{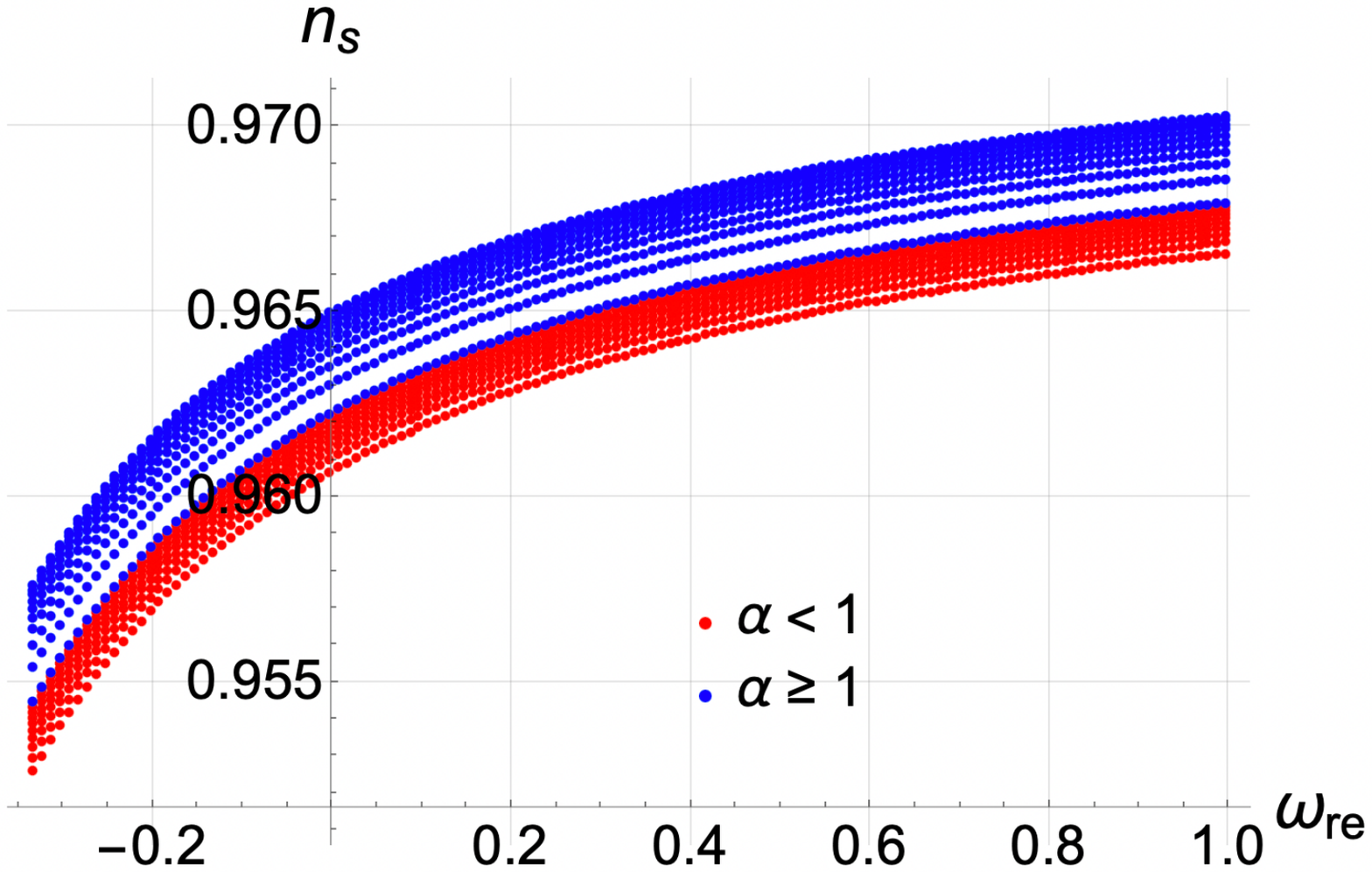}&
\includegraphics[width=3.in]{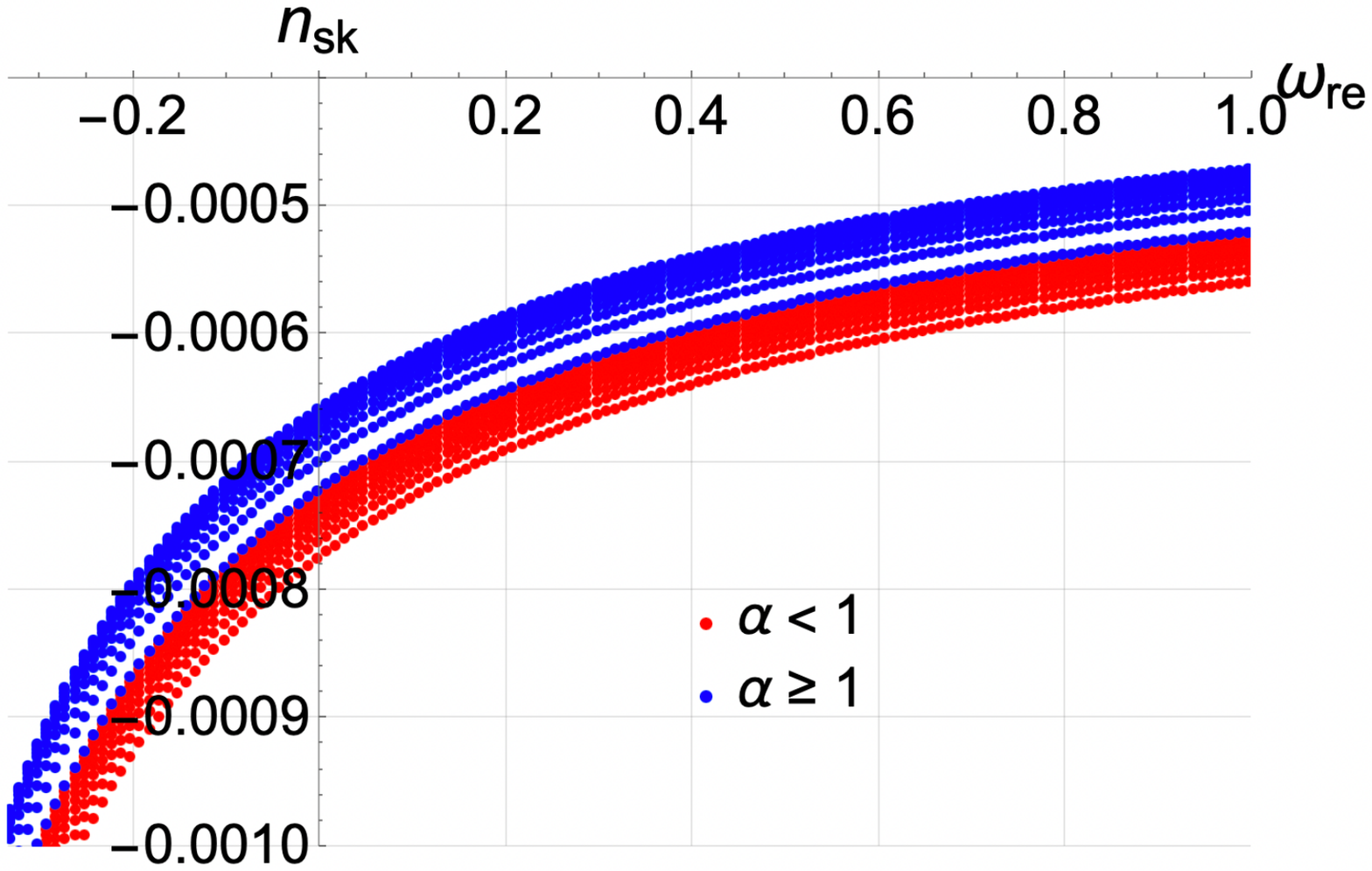}\\
\end{array}$
\caption{The plots show the scalar spectral index, $n_s$, and its running, denoted here by $n_{sk}$, as functions of the EoS parameter $\omega_{re}$ for the $\alpha$-Starobinsky model defined by eq.~(\ref{VS}). Shaded regions distinguish solutions with $\alpha < 1$ from those with $\alpha > 1$, according to the model parameter $\alpha$. Other details are the same as in Fig.~\ref{S_Tre_Nk}.}
\label{S_ns_nsk_wre}
\end{center}
\end{figure*}

\begin{figure*}[p]
\begin{center}
\includegraphics[width=4.in]{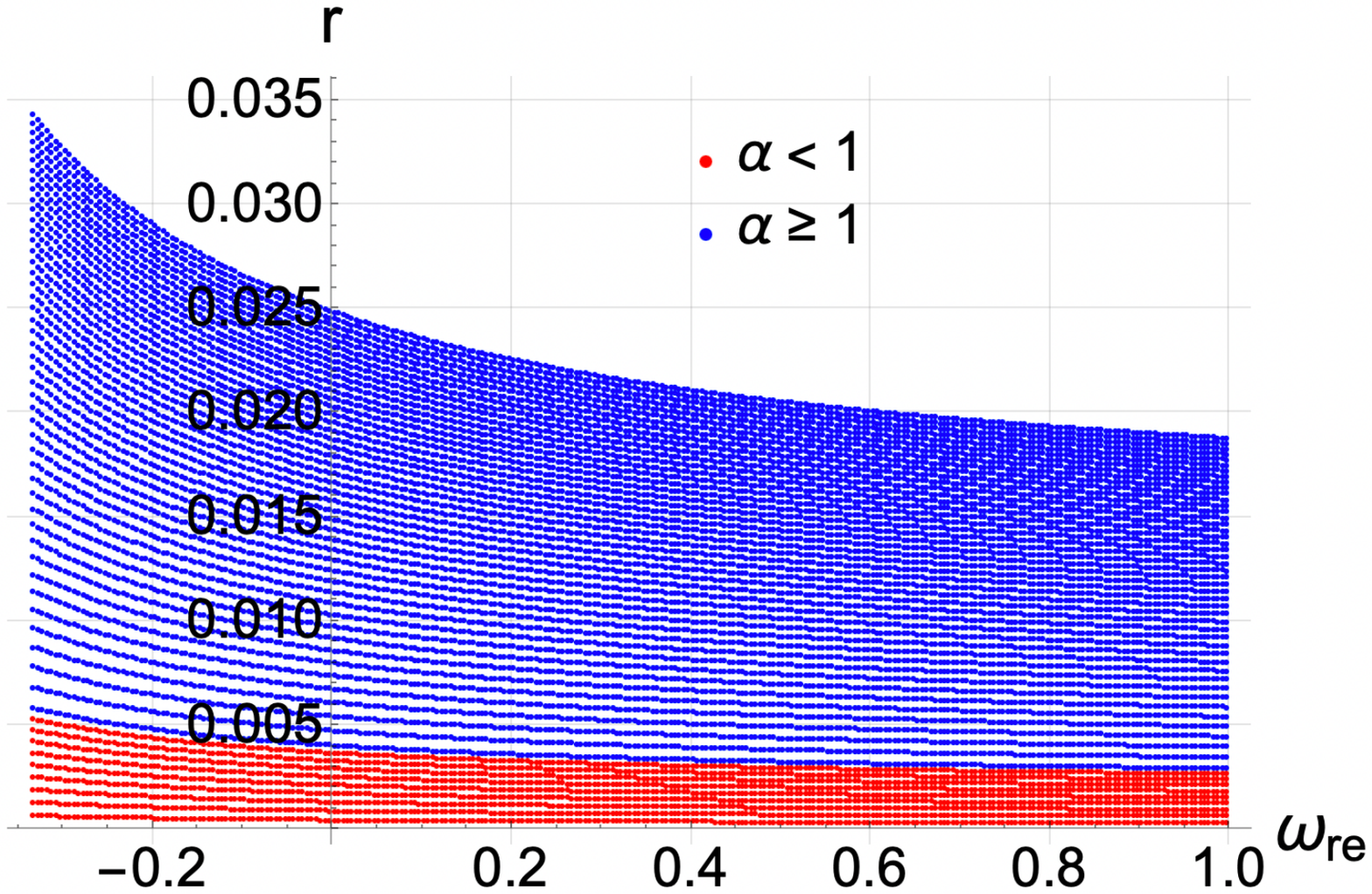}
\caption{The figure shows the tensor-to-scalar ratio, $r$, as a function of the EoS parameter $\omega_{re}$ for the $\alpha$-Starobinsky model defined by eq.~(\ref{VS}). Shaded regions differentiate solutions with $\alpha < 1$ from those with $\alpha > 1$, based on the model parameter $\alpha$. Other details are the same as in Fig.~\ref{S_Tre_Nk}.}
\label{S_r_wre}
\end{center}
\end{figure*}

\begin{figure*}[p]
\begin{center}
$\begin{array}{ccc}
\includegraphics[width=3.in]{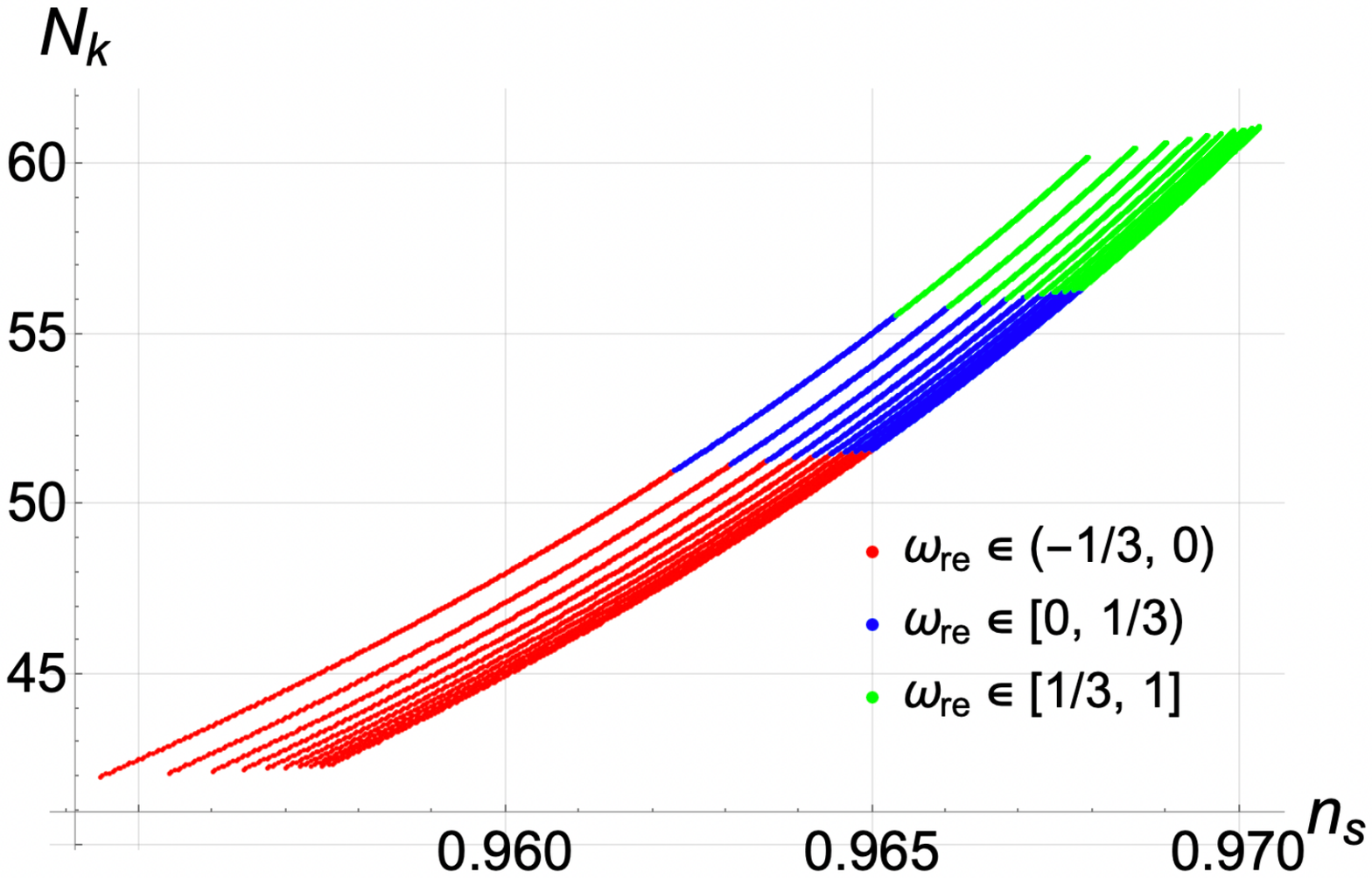}&
\includegraphics[width=3.in]{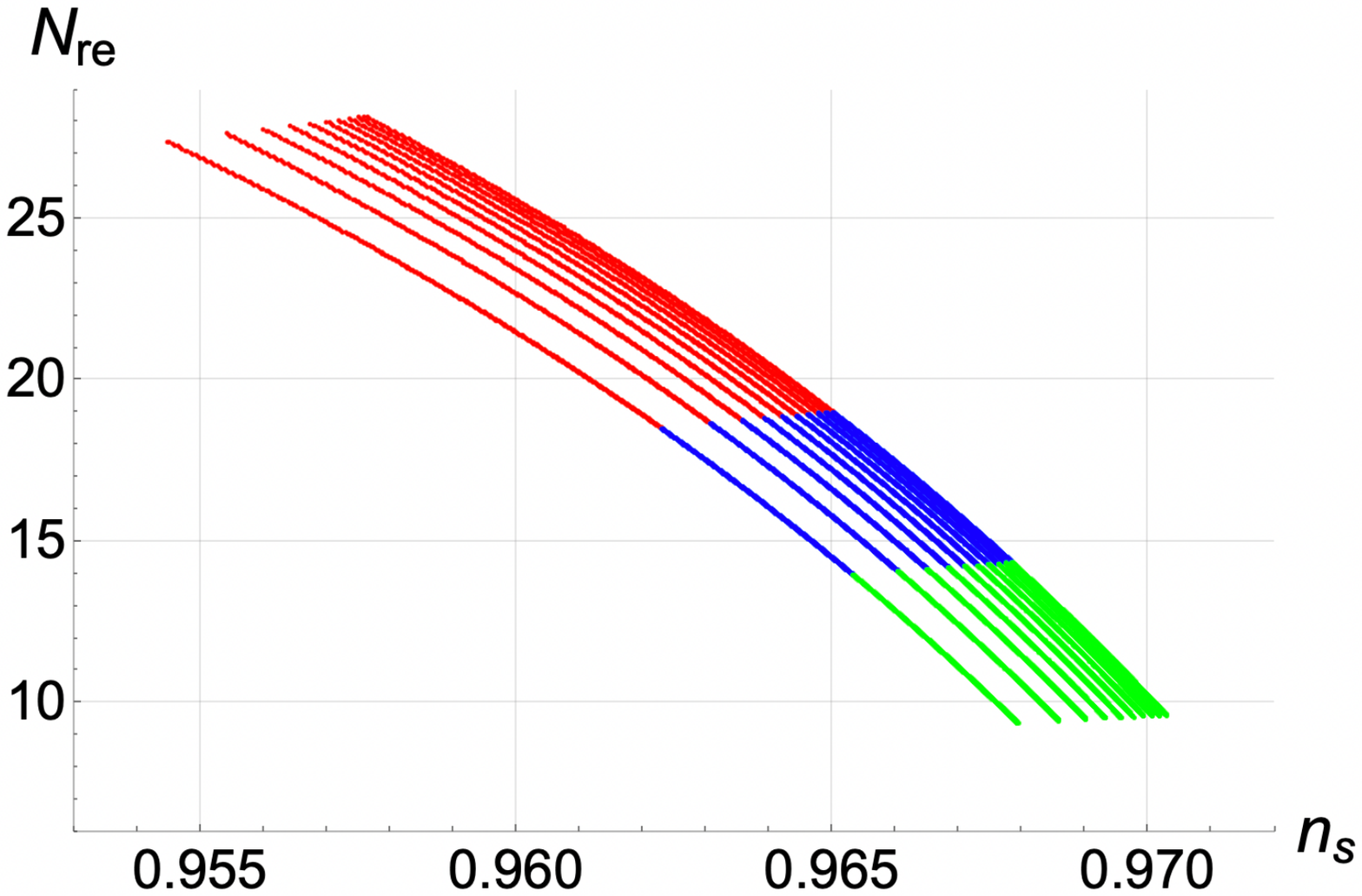}\\
\includegraphics[width=3.in]{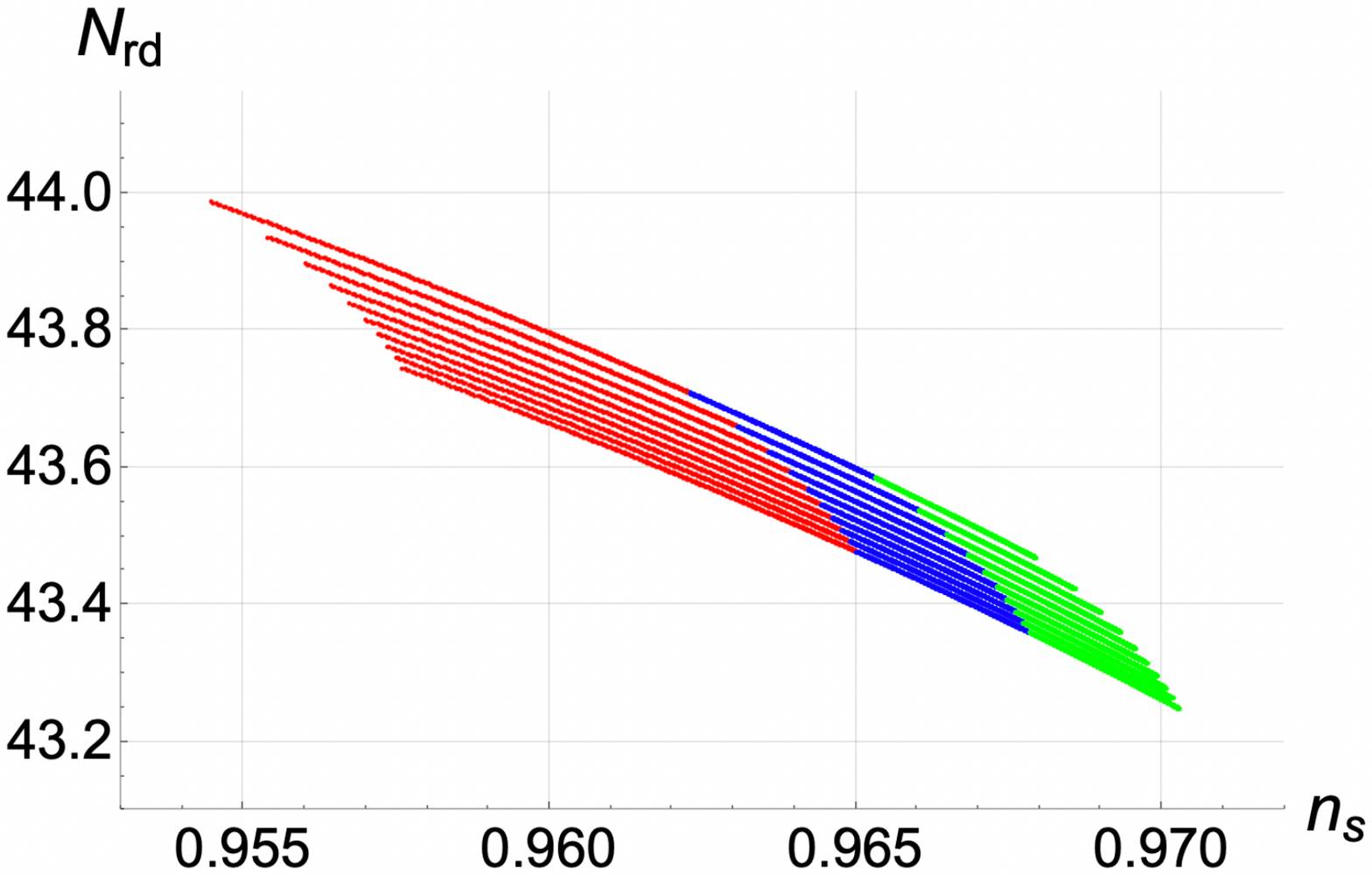}&
\includegraphics[width=3.in]{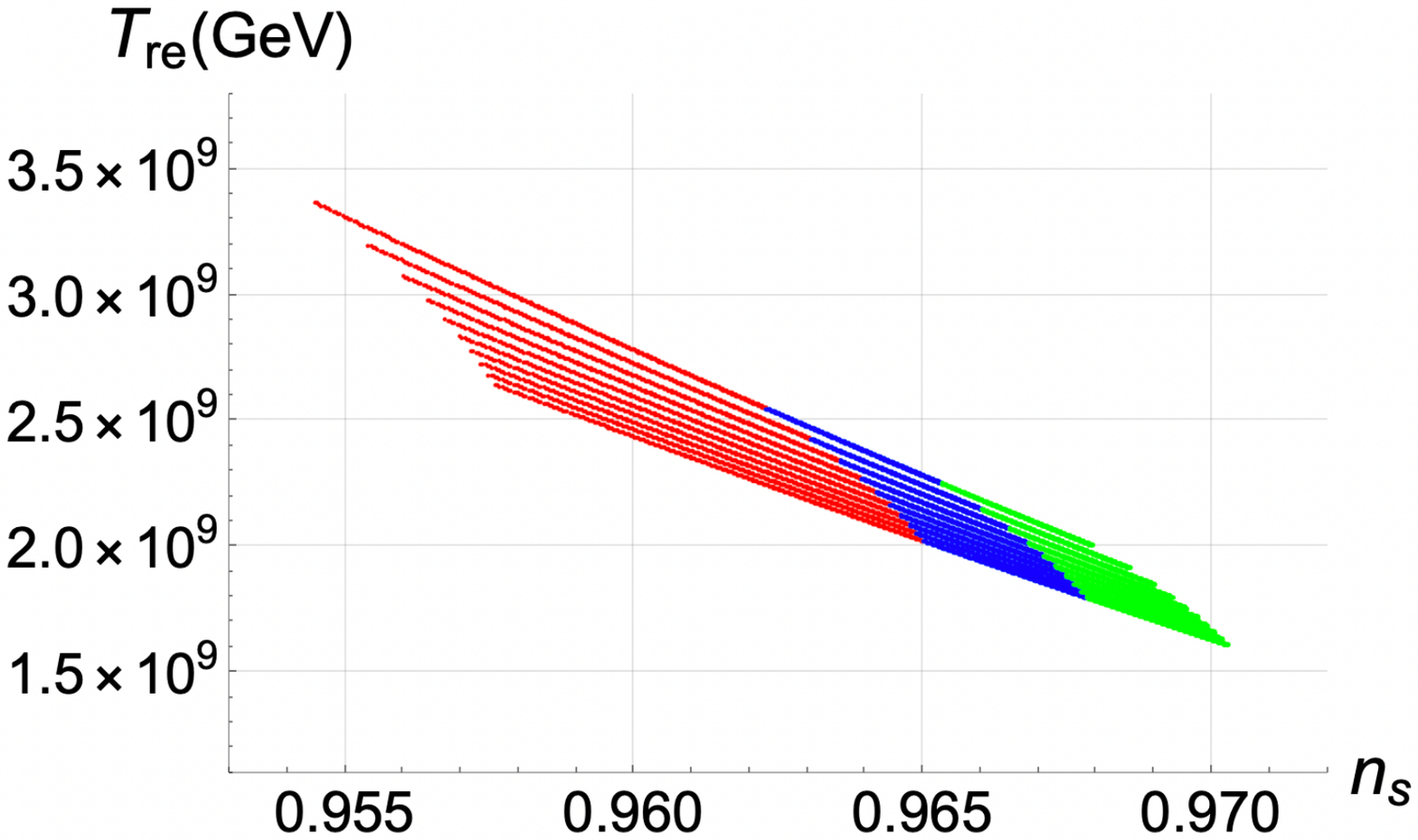} \\
\end{array}$
\caption{The plots show the number of $e$-folds during inflation, $N_k$, reheating, $N_{re}$, and during radiation, $N_{rd}$, as well as the reheating temperature at the end of reheating, $T_{re}$, as functions of the scalar spectral index, $n_s$, for the $\alpha$-Starobinsky model defined by eq.~(\ref{VS}). Details are the same as in Fig.~\ref{S_Tre_Nk}.}
\label{S_cosas_ns}
\end{center}
\end{figure*}

\begin{figure*}[p]
\begin{center}
$\begin{array}{ccc}
\includegraphics[width=3.in]{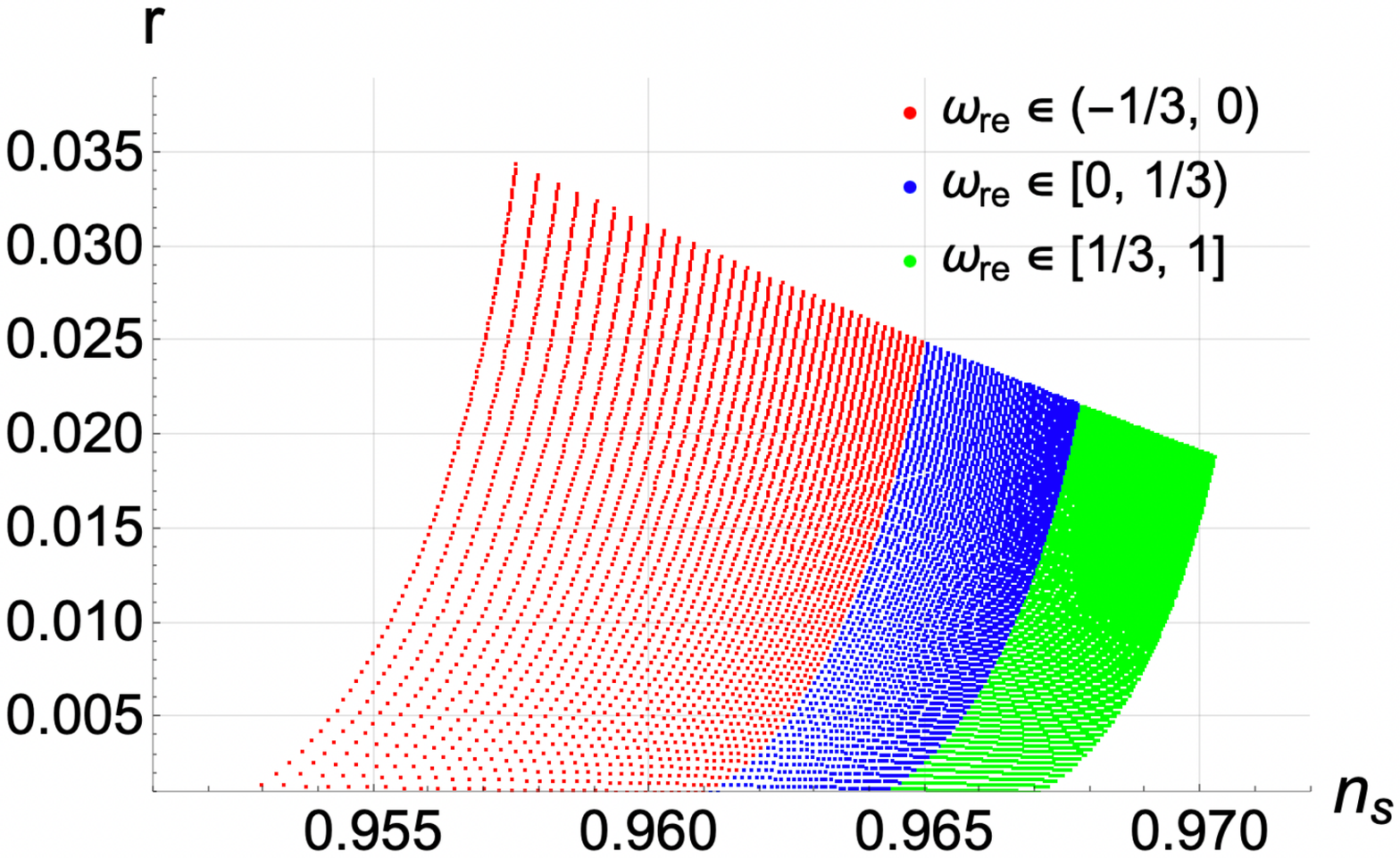}&
\includegraphics[width=3.in]{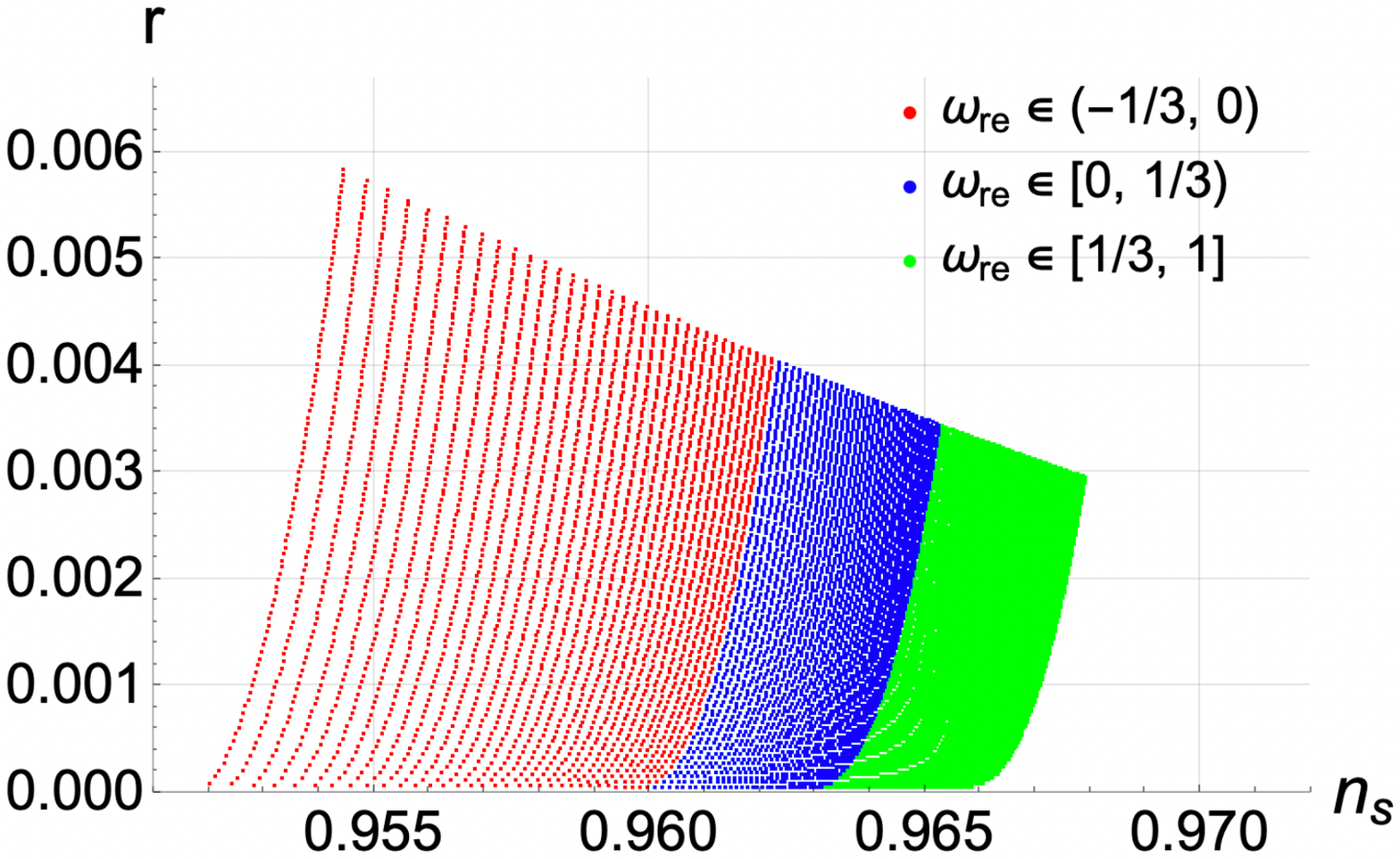}\\
\end{array}$
\caption{Plots of the tensor-to-scalar ratio, $r$, as a function of the scalar spectral index, $n_s$, for the $\alpha$-Starobisnky model defined by eq.~(\ref{VS}). The figure on the right-hand side corresponds to $\alpha \leq 1$ values of the model parameter. Other details are the same as in Fig.~\ref{S_Tre_Nk}.}
\label{S_rcosas_ns}
\end{center}
\end{figure*}

\begin{figure*}[p]
\begin{center}
\includegraphics[width=4.in]{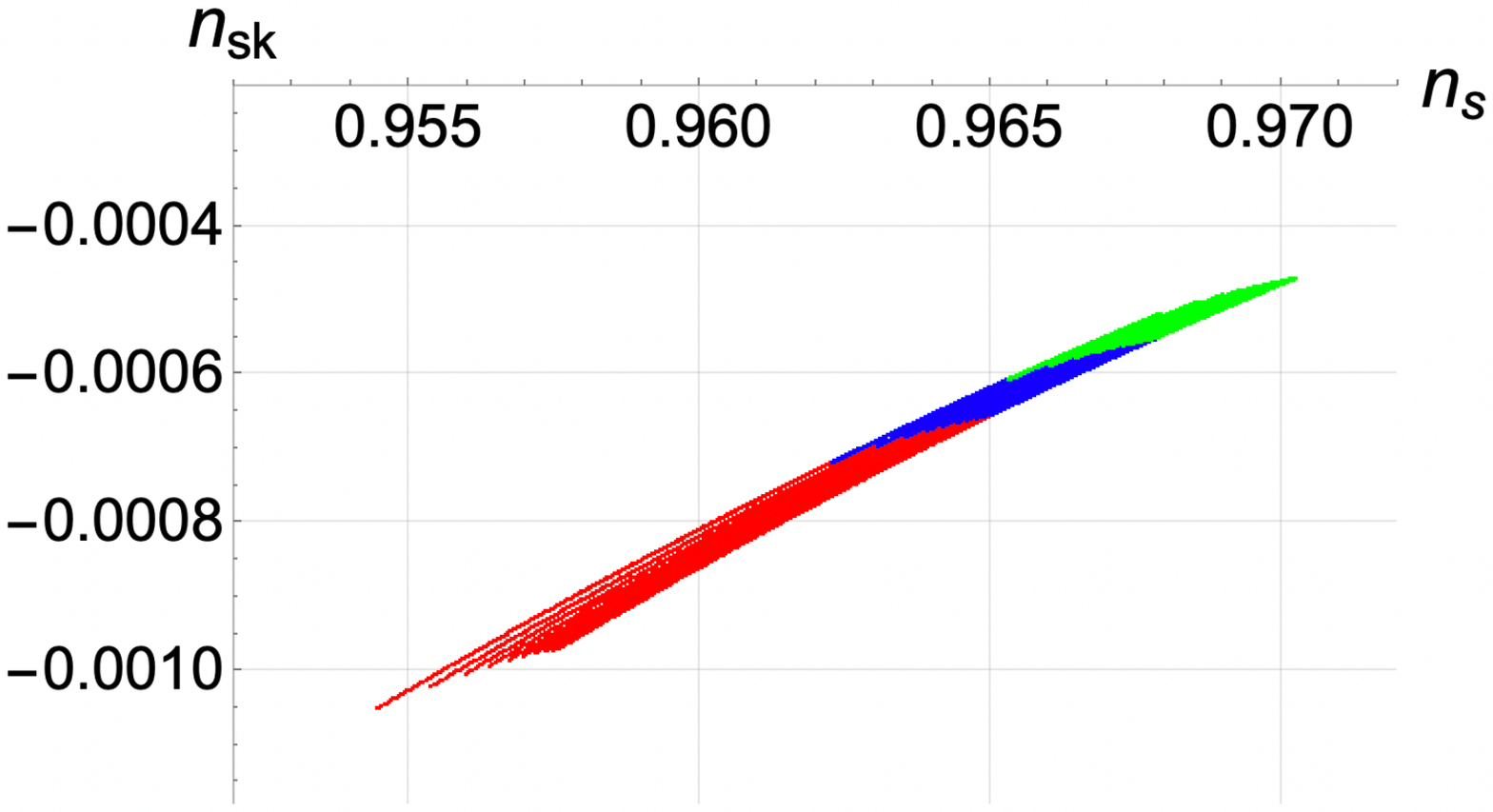}
\caption{Plot of the running of the scalar spectral index, denoted here by $n_{sk}$, as a function of $n_s$. for the $\alpha$-Starobinsky model defined by eq.~(\ref{VS}). Details are the same as in Fig.~\ref{S_Tre_Nk}.}
\label{S_nsk_ns}
\end{center}
\end{figure*}

The observables $n_s$ and $r$ can be easily determined in terms of $\phi_k$, as done for e.g., eq.~(\ref{OIns}) of the previous section. By eliminating $\phi_k$ from the expressions for these observables, one obtains
\begin{equation}
\label{alfaS}
\alpha = \frac{16r}{3\left(4\delta_{n_s} - r\right)^2},
\end{equation}
and the entire calculation process from Section \ref{Attractor} is followed. Table~\ref{S_bounds} presents the results of these calculations for the $\alpha$-Starobinsky model from eq.~(\ref{VS}), while Figures \ref{S_Tre_Nk} to \ref{S_nsk_ns} illustrate the evolution of various cosmological quantities, as described in their respective captions, with general details provided in Figure~\ref{S_Tre_Nk}.
\begin{table*}[htbp!]
 \begin{center}
{\begin{tabular}{cccc}
\small
Observable & range & others & range  \\
\hline \hline
{$\omega_{re}$} & $(-1/3,1)$ & $\alpha$ & $(3.2\times 10^{-17}, 153.5)$  \\
{$n_{s}$} & $(0.9578, 0.9711)$ & $N_{k}$ & $(42.0, 61.9)$  \\
{$r$} & $(1.7\times 10^{-19}, 0.068)$ & $N_{re}$ & $(29.0, 9.9)$ \\
{$n_{sk}$} & $(-0.00100,  -0.00048)$ & $N_{rd}$ & $(43.6, 42.8)$\\
{$n_{tk}$} & $(-0.00029,  -0.00018)$ & $T_{re}/\mathrm{GeV}$ & $(2.2\times 10^{9}, 1.0\times 10^{9})$  \\
\hline\hline
\end{tabular}}
\caption{Values of the parameter, observables, and cosmological quantities of interest for the $\alpha$-Starobinsky model defined by eq.~(\ref{VS}). A recently reported most probable value for vanilla inflation, $n_{sk} = -6.3 \times 10^{-4}$ \cite{Martin:2024nlo}, falls comfortably within the range above. Note how $T_{re}$ is tightly constrained. Details are the same as in Table~\ref{A_bounds}.}
\label{S_bounds}
\end{center}
\end{table*}
\newpage
\section{Discussion and Conclusions}\label{Con}
We have studied the connection between inflationary models and the reheating phase that follows inflation, with a dynamical reheating temperature, $T_{\text{re}}$, entering the equation that links these two phases. Our focus was on two widely-studied classes of models: generalized $\alpha$-attractor models and the $\alpha$-Starobinsky generalization, both motivated by supergravity and string theory. These models exhibit attractor behavior, ensuring that their predictions remain robust and consistent with cosmological observations. A crucial aspect of our work is the inclusion of an analytical expression for the reheating temperature, $T_{\text{re}}$, derived from inflaton decay. This expression plays a crucial role in determining important cosmological quantities, such as the scalar spectral index $n_s$ and the tensor-to-scalar ratio $r$. Interestingly, the maximum value of $T_{\text{re}}$ is of the order required to avoid overproduction of gravitinos during reheating, in the context of supergravity theories. Additionally, the lower bound for $T_{\text{re}}$ lies well above the usual limit imposed by nucleosynthesis constraints or any other arbitrary scale commonly assumed, keeping $T_{\text{re}}$ tightly constrained. By numerically solving the equation that links inflation to reheating, we explored how cosmological quantities and observables evolve as functions of the number of $e$-folds during inflation, $N_k$, the equation of state parameter during reheating, $\omega_{\text{re}}$, and the scalar spectral index $n_s$. Our results show a universal scaling behavior for $T_{\text{re}}$ in both the $\alpha$-attractor and $\alpha$-Starobinsky models, with $T_{\text{re}}$ scaling as a power of $1/N_k^{3/2}$ in the large-$N_k$ limit. This study complements previous Bayesian and numerical analyses by providing detailed numerical and analytical insights into the evolution of those quantities during inflation and the reheating phase. Our findings offer constraints on inflationary models based on observational data and emphasize the importance of reheating in shaping the early universe. In future work, these results could be refined by incorporating more general interactions between the inflaton and particle physics models, and by exploring their implications for the production of dark matter and other relics. Nevertheless, this analysis represents an important step towards a deeper understanding of the interplay between inflation and reheating.
\section*{Acknowledgments}

We would like to thank DGAPA-PAPIIT grant: IN110325 {\it Estudios en cosmolog\'ia inflacionaria, agujeros negros primordiales y
energ\'ia oscura}, UNAM, and the CONAHCYT ``SNII''  for its funding and support. 
\\
\\
{\bf Data Availability Statement}: No Data associated in the manuscript.

\end{document}